\documentclass{aastex63}

\usepackage{graphicx}	
\usepackage{amsmath}	
\usepackage{amssymb}	
\usepackage{wasysym}

\accepted{December 8, 2019}
\submitjournal{ApJ}

\shorttitle{Galactic archaeology of GRB hosts}
\shortauthors{Palla et al.}
\graphicspath{{./}}

\begin{document}

\title{Galactic archaeology at high redshift: inferring the nature of GRB host galaxies from abundances}

\correspondingauthor{Marco Palla}
\email{marco.ball94@gmail.com}

\author[0000-0002-3574-9578]{Marco Palla}
\affiliation{Dipartimento di Fisica, Sezione di Astronomia, Universit{\'a} degli Studi di Trieste, via G.B. Tiepolo 11, I-34131, Trieste, Italy}
\affiliation{IFPU - Institute  for  Fundamental Physics  of  the  Universe,  Via  Beirut  2,  I-34014,  Trieste,  Italy}

\author[0000-0001-7067-2302]{Francesca Matteucci}
\affiliation{Dipartimento di Fisica, Sezione di Astronomia, Universit{\'a} degli Studi di Trieste, via G.B. Tiepolo 11, I-34131, Trieste, Italy}
\affiliation{INAF, Osservatorio Astronomico di Trieste, via G. B. Tiepolo 11, I-34131, Trieste, Italy}
\affiliation{INFN, Sezione di Trieste, via A. Valerio 2, I-34100, Trieste, Italy}

\author[0000-0002-6175-0871]{Francesco Calura}
\affiliation{INAF, Osservatorio Astronomico di Bologna, via P. Gobetti 93/3, I-40129, Bologna, Italy}

\author[0000-0003-2501-2270]{Francesco Longo}
\affiliation{IFPU - Institute  for  Fundamental Physics  of  the  Universe,  Via  Beirut  2,  I-34014,  Trieste,  Italy}
\affiliation{INFN, Sezione di Trieste, via A. Valerio 2, I-34100, Trieste, Italy}
\affiliation{Dipartimento di Fisica, Universit{\'a} degli Studi di Trieste, via A. Valerio 2, I-34100, Trieste, Italy}



\begin{abstract}

We identify the nature of high redshift long Gamma-Ray Bursts (LGRBs) host galaxies by comparing the observed abundance ratios in the interstellar medium with detailed chemical evolution models accounting for the presence of dust.  We compare abundance data from long Gamma-Ray Bursts afterglow spectra to abundance patterns as predicted by our models for different galaxy types. We analyse $[X/Fe]$ abundance ratios (where $X$ is $C$, $N$, $O$, $Mg$, $Si$, $S$, $Ni$, $Zn$) as functions of $[Fe/H]$. Different galaxies (irregulars, spirals, spheroids) are, in fact, characterised by different star formation histories, which produce different $[X/Fe]$ vs. $[Fe/H]$ relations ($\lq\lq$time-delay model"). This allows us to identify the star formation history of the host galaxies and to infer their age (i.e. the time elapsed from the beginning of star formation) at the time of the GRB events. Unlike previous works, we use newer models in which we adopt updated stellar yields and prescriptions for dust production, accretion and destruction. We consider a sample of seven LGRB host galaxies. Our results suggest that two of them (GRB  050820, GRB 120815A) are star forming spheroids, two (GRB 081008, GRB 161023A) are spirals and three (GRB 090926A, GRB 050730, GRB 120327A) are irregulars. 
The inferred ages of the considered host galaxies span from $10$ $Myr$ to slightly more than $1$ $Gyr$.

\end{abstract}






\section{Introduction} \label{s:intro}

Gamma-Ray Bursts (GRBs) are sudden and extremely powerful flashes of gamma radiation. They originate at cosmological distances and their typical duration varies from a few milliseconds to $\sim~10^3$ seconds. GRBs lasting more than 2 seconds are generally called long GRBs (LGRBs, \citealt{Kouveliotou93}). LGRBs are often associated to the death of very massive stars, in particular core-collapse supernovae (CC-SNe) of Type Ib and Ic. In fact,  the study of GRB afterglows, namely the prolonged, lower energy radiation (i.e. in the X-ray, optical, and radio band) visible for several days after the GRB event, has clearly indicated the association of 27 LGRBs (until 2016, \citealt{Hjorth16}) with such kinds of SNe.\\
In this scenario, with the constantly increasing number of high redshift GRBs analysed so far, the afterglow spectra can be used to probe the type of star-forming galaxies, in particular those at high redshift. 
Furthermore, the understanding of the nature of GRB host galaxies can give stringent constraints on GRB progenitor models, favouring the single progenitor (collapsar by \citealt{MacFadyen99,Woosley06} or millisecond magnetar by \citealt{Wheeler00,Bucciantini09}) or the binary progenitor models \citep{Fryer05,Demeters08,Podsiadlowski10}. Many attempts have been made in the past to characterise GRB host galaxies \citep{LeFloch03,Fruchter06,Savaglio09,Levesque10,Boissier13, Schulze15,Kruhler15,Perley16b,Arabsalmani18}. Different studies (e.g. \citealt{Vergani15,Perley16b,Palmerio19}) have  demonstrated that LGRBs prefer subsolar metallicity host galaxies. This means that, with respect of the standard star forming galaxy population, at lower redshift there is a bias towards dwarf and subsolar metallicity galaxies \citep{Fruchter06,Vergani15}. The bias is not necessairily expected to occur also at high redshift, where most galaxies of all types are suffering the first episodes of star formation and have still sub-solar metallicities.\\ 
Following the idea developed by \citet{Calura09} and adopted also by \citet{Grieco14}, in this paper we use chemical evolution models for different galactic morphological types (irregular, spiral, spheroids) which predict the evolution of the abundances of the main chemical elements ($H$, $He$, $C$, $N$,$\alpha$-elements\footnote{elements synthetised by capture of $\alpha$ particles. Examples are $O$, $Mg$, $Si$, $S$.}, $Fe$, $Ni$, $Zn$, etc.), to identify the nature of GRB host galaxies. The basic idea beneath this procedure derives from the $\lq\lq$time-delay model" \citep{Matteucci01,Matteucci12}, which explains the observed behaviour of $[X/Fe]$\footnote{by definition: $[X/Y] = \log(X/Y) - \log(X_\odot/Y_\odot)$, where $X$, $Y$ are abundances in mass in the ISM for the object studied and $X_\odot$, $Y_\odot$ are solar abundances in mass.} vs. $[Fe/H]$ , with $X$ being any chemical element, as due to the different roles played by CC and Type Ia SNe (white dwarfs exploding in binary systems) in the galactic chemical enrichment process. Based on the fact that the [$\alpha/Fe$] ratio evolution is predicted to be quite different in different star formation (SF) regimes \citep{Matteucci90,Matteucci01}, galaxies of different morphological type show a different behaviour of the $[\alpha/Fe]$ vs. $[Fe/H]$ abundance plot.\\
The models we are adopting for different galaxy types differ mainly by the star formation history and take into account possible condensation of the main metals ($C$, $\alpha$-elements, $Fe$, $Ni$) into dust. Observations of LGRBs in mid-IR and radio bands show clearly the presence of dusty environments in many hosts (e.g. \citeauthor{Perley09} \citeyear{Perley09}, \citeyear{Perley13}, \citeyear{Perley17}; \citealt{Greiner11,Hatsukade12,Hunt14}). Therefore, to understand the nature of the hosts on the basis of gas abundance data, it is fundamental to consider dust in galactic evolutionary models. 
With respect to the previous works of \citet{Calura09} and \citet{Grieco14}, based on chemical evolution models with dust by \citet{Calura08}, in this paper we adopt a much improved formulation of dust evolution. In particular, we use newer and more accurate prescriptions for dust production \citep{Piovan11,Gioannini17a} and other dust processes in the ISM, such as accretion and destruction \citep{Asano13}, as well as for the stellar yields \citep{Karakas10,Doherty14a,Doherty14b,Nomoto13}, which are already tested on the solar neighbourhood abundances. With respect to the host galaxies, we consider five afterglow spectra already studied by \citet{Calura09} and \citet{Grieco14}, plus a couple of systems never considered before in such analyses.\\ 
The paper is organised as follows: Section \ref{s:sample} shows and briefly explains the observational data considered in this work. Section \ref{s:chem_model} describes the chemical evolution models, specifying also the dust prescriptions (production, accretion and destruction) used throughout this work. In Section \ref{s:results} we explain the parameter values in the adopted models and we show 
the results derived from the comparison between the model predictions and the abundance data of the analysed hosts. Finally, in Section \ref{s:conclusion} some conclusions are drawn. 


\section{Host galaxy sample} \label{s:sample}

In order to constrain the nature of GRB host galaxies we have chosen GRBs with a quite large number of observed abundances measured in their environment: GRB 050730, GRB 050820 \citep{Prochaska07}, GRB 081008 \citep{DElia11}, GRB 090926A \citep{DElia10}, GRB 120327A \citep{DElia14}, GRB 120815A \citep{Kruhler13}, GRB 161023A \citep{deUgarte18}. In Table \ref{t:observation} the observational data (redshift, stellar mass, abundance ratios) are shown for each of the studied hosts.\\
We decided to include in our sample data already used in the previous GRB host identification works of \citet{Calura09} (GRB 050730, GRB 050820) and \citet{Grieco14} (GRB 081008, GRB 120327A, GRB 120815A). The main reason of their inclusion is to test the results obtained with older chemical evolution models \citep{Calura08} containing less updated stellar yields and dust prescriptions (see Section \ref{s:intro}). In this way, we can see if newer models lead to different results with respect to older ones, highlighting the importance of using more accurate models to reach more robust conclusions.\\
In Table \ref{t:observation} we do not present $[C/Fe]$ and $[O/Fe]$ ratios, although they are available in almost all the studies considered. This decision was made because they are lower/upper limits or abundances affected by biases (lines saturation, blending) in their determination.

\begin{table*} 
\tablenum{1}
\flushleft
\caption{Physical characteristics and abundances of the hosts of the GRBs considered here.}
\begin{tabular}{c | *{7}c}
\hline
 & GRB 050730 & GRB 050820 & GRB 081008 & GRB 090926A & GRB 120327A & GRB 120815A & GRB 161023A \\[0.1cm]
 \hline
$z$ & $3.969$ & $2.165$ & $1.968$ & $2.107$ & $2.815$ & $2.360$ & $2.710$\\[0.05cm]
$M_*$ & $<10^{9.46} M_\odot$ & $10^9 M_\odot$ & $<10^{9.18} M_\odot$ & - & - & $10^{10} M_\odot$ & $<10^{9.6} M_\odot$ \\[0.1cm]
\hline
$[Fe/H]$ & $-2.59\pm0.10$ & $-1.69\pm0.10$ & $-1.19\pm0.11$ & $-2.29\pm0.09$ & $-1.73\pm0.10$ & $-2.18\pm0.11$ & $1.81\pm0.04$\\
$[N/Fe]$ & $-0.47\pm0.14$ & $>0.44$ & - & $-0.97\pm0.08^{1}$ & $0.28\pm0.15$ & - & -  \\ 
$[Mg/Fe]$ & $<0.88$ & $0.91\pm0.14$ & - & $>-0.83\pm0.12$ & $0.46\pm0.14$ & - & $0.32\pm0.20$\\
$[Si/Fe]$ & $>-0.23$ & $>0.50$ & $0.32\pm0.15$ & $-0.12\pm0.12$ & $0.61\pm0.15$ & $\apprge1.02\pm0.22$ & $0.37\pm0.06$\\
$[S/Fe]$ & $0.35\pm0.14$ & $1.08\pm0.14$ & - & $0.34\pm0.13$ & $0.34\pm0.13$ & $\apprle 1.31\pm0.28$ & $0.66\pm0.06$\\
$[Ni/Fe]$ & $-0.06\pm0.14$ & $0.16\pm0.14$ & $-0.10\pm0.16$ & $0.28\pm0.16$ & $0.10\pm0.13$ & $0.22\pm0.16$ & - \\
$[Zn/Fe]$ & - & $1.04\pm0.14$ & $0.67\pm0.15$ & - & $0.56\pm0.15$ & $1.10\pm0.15$ & $0.70\pm0.08$\\
\hline
\end{tabular}
\tablecomments{References for the data in Section \ref{s:sample}. All abundance ratios are normalised to \citet{Asplund09} solar abundances. For GRB 050820, $M_*$ comes from \citet{Kruhler17}, whereas for GRB 050730, 081008 comes from \citet{Perley16b}.}
$^{1}$\footnotesize{not taken into account in the analysis (see \ref{sss:090926A})}
\label{t:observation}
\end{table*}

\section{Chemical evolution models including dust}  \label{s:chem_model}

We trace the evolution of chemical abundances in galaxies of different morphological types (spheroids, spirals, irregulars) by means of chemical evolution models including dust evolution. These models relax the instantaneous recycling approximation (IRA), taking into account stellar lifetimes. All the models assume that galaxies formed by primordial gas infall which accumulates into a preexisting dark matter halo on different timescales and evolve suffering galactic winds. In our work, we do not consider galaxy interactions.  
Several chemical evolution papers have demonstrated that interactions can be simulated by enriched gas infall from a companion galaxy \citep{Spitoni15,Spitoni16}.
In \citet{Spitoni15} is shown that external, primordial infall is more important than enriched one, especially in the early phases of galaxy evolution.

\subsection{The Birthrate Function}     \label{ss:birthrate_f}
A fundamental parameter for these models is the birthrate function $B(m,t)$, which represents the number of stars formed in the mass interval $[m, m + dm]$ and in the time interval $[t, t + dt]$. It 
is expressed as the product of two independent functions:
\begin{equation}
   B(m, t) = \psi(t)\phi(m),
\end{equation}
where the term $\psi(t)$ is the star formation rate (SFR), while $\phi(m)$ represents the initial mass function (IMF).\\
The SFR is the rate at which stars form per unit time and it is generally expressed in units of $M_\odot yr^{-1}$. To parametrise the SFR, in our models we adopt the Schmidt-Kennicutt law \citep{Schmidt59,Kennicutt98}:
\begin{equation}
    \psi(t) = \nu G(t)^k.
\label{e:SFR}
\end{equation}
In this expression, $\nu$ is the star formation efficiency, namely the inverse of the time scale of star formation (expressed in $Gyr^{-1}$), which varies depending on the morphological type of the galaxy (see Tables \ref{t:models1}, \ref{t:models2}). In particular, the variation of $\nu$ determines the different SFR in galaxies of different morphological type, decreasing from spheroids to spirals and then irregulars. $G(t) = M_{ISM} (t)/M_{inf}$ is the ISM mass fraction relative to the infall mass. This latter is the total gas mass fallen into the dark matter halo during the evolution history;  therefore, the final stellar mass will always be lower than the infall mass because of the galactic winds. The parameter $k$ is set equal to $1$.\\
The IMF represents the mass distribution of stars at their birth. It is assumed to be constant in space and time and normalised to unity in the mass interval $[0.1M_\odot , 100M_\odot ]$.\\
We start by assuming a \citet{Salpeter55} IMF for all galaxies:
\begin{equation}
    \phi_{Salp}(m) \propto m^{-(1+1.35)}.
\end{equation}
For spheroidal galaxies, the computations are done also with a top-heavy single-slope IMF,
\begin{equation}
    \phi_{top}(m) \propto m^{-(1+0.95)},
    \label{e:top_IMF}
\end{equation}
since in more massive spheroids an overabundance of massive stars at early times is necessary to explain many of their features, such as the colour-luminosity relation \citep{Gibson97} and the observed $[\alpha/Fe]$ (e.g. \citealt{DeMasi18_1}) in local galaxies and the isotopic ratios \citep{Zhang18} in high redshift ones.\\
For spirals and irregulars instead, in addition to the \citet{Salpeter55} IMF, we use also a \citet{Scalo86} IMF, derived for the solar vicinity:
\begin{gather}
\phi_{Scalo}(m) \propto \bigg \{\begin{array}{rl}
					0.19\cdot m^{-(1+1.35)} & m\leq 2 M_\odot \\
					0.24\cdot m^{-(1+1.7)} \hspace*{0.1cm}    & m > 2 M_\odot,\\
				  \end{array}
\end{gather}
which in general fits better the features of spiral disks (in particular the Milky Way disk) than the \citet{Salpeter55} one \citep{Chiappini01,Romano05}.

\subsection{Chemical Evolution Equations}   \label{ss:chem_eq}

We follow the evolution of the abundance of a given element in the gas by means of the following equation:
\begin{equation}
\dot{G}_i(t)=-\psi(t)X_i(t)+ R_i(t) + \dot{G}_{i,inf}(t) - \dot{G}_{i,w}(t),
\label{e:chem_evo}
\end{equation}
where $G_i(t)=G(t)X_i(t)$ is the mass of the element $i$ in the ISM normalised to the infall mass and $X_i(t)$ represents the fraction of the element $i$ in the ISM at a certain time $t$.\\
The four terms on the right side are:
\begin{enumerate}
    \item $-\psi(t)X_i (t)$ represents the rate at which the element $i$ is removed from the ISM due to the star formation process.
    \item $R_i (t)$ is the rate at which the element $i$ is restored into the ISM from stars thanks to SN explosions and stellar winds. Inside this term the nucleosynthesis prescriptions of the specific element $i$ are taken into account (see \ref{sss:nucleosynthesis}). In order to relax the IRA, $R_i (t)$ has the following form, as shown by \citet{MatteucciGreggio86}:
\begin{multline}
\hspace{4.7cm}R_i(t)=\int_{M_L}^{M_{Bm}}\!{\psi(t-\tau_m)Q_{mi}(t-\tau_m)\phi(m)}\,dm +\\
+A\int_{M_{Bm}}^{M_{BM}}\phi(m) \bigg[ \int_{\mu_{min}}^{0.5} f(\mu)	\psi(t-\tau_{m2})Q_{mi}(t-\tau_{m2})\,d\mu \bigg] \,dm +\\
+ (1-A)\int_{M_{Bm}}^{M_{BM}}\psi(t-\tau_m)Q_{mi}(t-\tau_m)\phi(m)\,dm +\\ +\int_{M_{BM}}^{M_{U}}\psi(t-\tau_m)Q_{mi}(t-\tau_m)\phi(m)\,dm . \\	
\label{e:restitution_stars}
\end{multline}
The first integral is the rate at which an element $i$ is restored into the ISM by single stars with masses in the range $[M_L,M_{B_m}]$, where $M_L$ is the minimum mass at a certain time $t$ contributing to chemical enrichment (for $t_f=14Gyr$, $M_L\sim0.8M_\odot$ ) and $M_{B_m}$ is  the  minimum mass for a binary system giving rise to a Type Ia SN ($M_{B_m}=3M_\odot$). The quantities $Q_{mi}(t-\tau_m)$, where $\tau_m$ is the lifetime of a star of mass $m$, contain all the information about stellar nucleosynthesis for elements either produced or destroyed or ejected without been processed \citep{Talbot71}.\\
The second term represents the material restored by binaries, with masses between $M_{B_m}$ and $M_{B_M}=16M_\odot$, which explode as Type Ia SNe. For these SNe a single degenerate scenario (SD) is assumed, where a $C$-$O$ white dwarf explodes after it exceeds the Chandrasekar mass ($1.44M_\odot$). $A$ is the parameter representing the fraction of binary systems able to produce a Type Ia SN and its value is set to reproduce the observed rate of Type Ia SNe. In this term both $\psi$ and $Q_{mi}$ refer to the time $t-\tau_{m2}$ where $\tau_{m2}$ indicates the lifetime of the secondary star of the binary system, which regulates the explosion timescale. $\mu=m_2/m_B$ is the ratio between the mass of the secondary component ($m_2$) and the total mass of the binary ($m_B$) and $f(\mu)$ represent the distribution function of this ratio.\\
The third integral represents the contribution given by single stars lying in the mass range [$M_{B_m} , M_{B_M}]$ which do not produce type Ia SNe events. If the mass $m > 8M_\odot$ , they explode as e-capture or CC-SNe, otherwise they die as white dwarfs.\\
The last term of \eqref{e:restitution_stars} refers to the material recycled back to the ISM by stars more massive than $M_{B_M}$, i.e. by the high mass CC-SNe up to $M_U=100M_\odot$.
We do not consider explicitly the effects of close massive binaries on the galactic chemical enrichment. Previous works have shown that the stellar yields are not substantially affected by them (see \citealt{DeDonder02}). 
  
    \item $\dot{G}_{i,inf}(t)$ represents the rate of infall of gas of the $i$-th element in the system. It is expressed as:
    \begin{equation}
\dot{G}_{i,inf}(t) \propto X_{i,inf} e^{-t/\tau_{inf}},
\label{e:infall_rate}
\end{equation}
Here $X_{i,inf}$ is the fraction of the element $i$ in the infalling gas, which has primordial composition in the models. The quantity $\tau_{inf}$ is the infall timescale, defined as the characteristic time at which half of the total mass of the galaxy has assembled and is set to satisfy observational constraints for the studied galaxies. This parameter varies with galactic type, increasing from spheroids to spirals and irregulars, in agreement with previous works (e.g. \citealt{Gioannini17b}).
    \item The last term of Equation \eqref{e:chem_evo} represents the outflow rate of the element $i$ due to galactic winds (GWs), developing when the thermal energy of the gas (heated by SNe explosions and stellar winds) exceeds its binding energy (for details see \citealt{Bradamante98}). The outflow rate has the form:
    \begin{equation}
\dot{G}_{i,w}(t)= \omega_i \psi(t),
\end{equation}
where $\omega_i$ is the wind rate parameter (i.e. mass loading factor) for the element $i$, a free parameter tuned to maximise the agreement with the observed galaxy features. In our models we do not use differential winds, so $\omega_i$ will be the same for all elements.
\end{enumerate}

\subsubsection{Nucleosynthesis Prescriptions} \label{sss:nucleosynthesis}

We compute in detail the contribution to chemical enrichment of the ISM of low-intermediate mass stars (LIMS), Type Ia and CC-SNe (Type II, Ib/c). To do this, we adopt specific stellar yields for all these stars. The yields are the amount of both newly formed and pre-existing elements injected into the ISM by dying stars.\\
In this paper we adopt mass and metallicity dependent stellar yields:
\begin{itemize}
    \item for LIMS ($0.8M_\odot<m<9M_\odot$) we use yields by \citet{Karakas10} for stars with mass lower than $6M_\odot$, while for super-AGB (SAGB) stars and e-capture SNe, with mass between $6M_\odot$ and $9M_\odot$ , we use yields by  \citeauthor{Doherty14a}(\citeyear{Doherty14a}, \citeyear{Doherty14b}).
    \item For massive stars, that explode as CC-SNe ($m>9M_\odot$), we adopt yields by \citet{Nomoto13} (revised version of \citealt{Koba06}). For nitrogen, calculation are performed also with the prescriptions by \citet{Matteucci86}.
    \item For Type Ia SNe we use the yields by \citet{Iwamoto99}.
\end{itemize}

\subsection{Dust Evolution Equation}    \label{ss:dust_eq}

Here we adopt the same formalism used in previous works of chemical evolution with dust \citep{Dwek98,Calura08,Gioannini17a,Vladilo18}. The equation governing the dust evolution is quite similar to \eqref{e:chem_evo}, but it includes other terms describing dust processes in the ISM. For a given element $i$, we have:
\begin{equation}
\begin{split}
\dot{G}_{i,dust}= - \psi(t) X_{i,dust}(t) + R_{i,dust}(t) + \dot{G}_{i,dust,accr}(t) +\\
-\dot{G}_{i,dust,destr}(t) - \dot{G}_{i,dust,w}(t), \hspace*{1.5cm}
\end{split}
\label{e:dust_evo}
\end{equation}
where $G_{i,dust}(t)$ and $X_{i,dust}(t)$ are the same of Equation \eqref{e:chem_evo} but only for the dust phase.\\
The five terms on the right side of Equation \eqref{e:dust_evo} are the following:
\begin{enumerate}
    \item The first term concerns the rate of dust astration. In other words, this is the process of removal of dust from the ISM due to star formation.
    \item $R_{i,dust}(t)$, similarly to $R_i(t)$ for Equation \eqref{e:chem_evo}, is the rate at which the element $i$ is ejected into the ISM in the form of dust by star. The term is also called dust production rate (DPR).
    \item The third term is the dust accretion rate (DAR) for the element $i$, which is the rate of dust mass enhancement due to grain growth by accretion processes in the ISM.
    \item $\dot{G}_{i,dust,destr}(t)$ is the dust destruction rate (DDR) for the $i$-th element, namely the rate at which dust is destroyed by SN shocks.
    \item The last term of Equation \eqref{e:dust_evo} indicates the rate of dust, in the form of element $i$, expelled by GWs. In the model we assume that dust and gas in the ISM are coupled, so the wind parameters are the same for the elements in the gas and dust phases.
\end{enumerate}
In the next paragraphs we will discuss the second, third and fourth term in more detail.

\subsubsection{Dust Production}     \label{ss:d_prod}
The interstellar dust is produced by stars: depending
on the physical structure of the progenitor (type of star, mass, metallicity), different amounts of dust species can originate.\\
We can summarise the second term of Equation \eqref{e:dust_evo} in this way (the complete expression can be found in \citealt{Gioannini17a}):
\begin{equation}
    R_{i,dust}(t)=\delta_i^{AGB}R_i^{LIMS}(t)+\delta_i^{CC}R_i^{CC-SN}(t).
\end{equation}
The terms $\delta_i^{AGB}$, $\delta_i^{CC}$ are the condensation efficiencies and represent the fraction of the element $i$ expelled by stars (AGB and CC-SNe, respectively) which goes into the ISM in the dust phase.\\
Following \citet{Gioannini17a}, the dust sources considered in this work are:
\begin{itemize}
    \item AGB (LIMS): in LIMS, the cold envelope during the AGB phase is the best environment in which nucleation and formation of dust seeds can occur, since previous phases do not present favourable conditions (small amount of ejected material, wind physical conditions) for producing dust. In the dust production process, stellar mass and metallicity play a key role in determining the dust species formed: this happens because $m$ and $Z$ are crucial to set the number of thermal pulses occurring, which define the surface composition of the star (e.g. \citealt{Ferrarotti06}; 
\citealt{DellAgli17}).\\
    In this paper we adopt the condensation efficiencies, dependent both on mass and metallicity, computed by \citet{Piovan11}, already presented and used in \citet{Gioannini17a}.
    \item CC-SNe: this is the other fundamental source of dust besides AGB stars. Evidence of dust presence in historical supernova remnants, such as SN1987A, Cas A, Crab Nebula, were observed (\citealt{Gomez13} and references therein). In particular from SN1987A observations, we now know that this SN produced up to $0.7M_\odot$ of dust \citep{Danziger91}. Despite of this, the picture is far from being totally clear. This is due to the uncertainties about the amount of dust destroyed by the reverse shock produced by the explosion of a SN (see \citealt{Gioannini17a} for a more detailed discussion).\\
    Also in this case we adopt the condensation efficiencies provided by \citet{Piovan11}, presented and used in \citet{Gioannini17a}. These $\delta_i^{CC}$ take into account both the processes of dust production and destruction by CC-SNe, but most importantly give us the possibility to choose between three different scenarios for the surrounding environment: low density ($n_H=0.1$ $cm^{-3}$), intermediate density ($n_H=1$ $cm^{-3}$) and high density ($n_H=10$ $cm^{-3}$). The higher is the density, the higher is the resistance that the shock will encounter, and the higher will be the dust destroyed by this shock. Between the three possibilities, in this work we adopt only $\delta_i^{CC}$ for $n_H=0.1$ $cm^{-3}$ and $n_H=1$ $cm^{-3}$. We make this choice following \citet{Gioannini17b}, where the intermediate density scenario reproduces the amount of dust detected in some high redshift starbursts as well as the dust-to-gas ratio (DGR) in spirals of the KINGFISH survey (\citealt{Kennicutt11}), whereas low density $\delta_i^{CC}$ are more indicated to explain the DGRs observed in the Dwarf Galaxy Survey (\citealt{Madden13}).
\end{itemize}
In this work, we assume that Type Ia SNe do not produce any dust, as in \citeauthor{Gioannini17a} (\citeyear{Gioannini17a}, \citeyear{Gioannini17b}).

\subsubsection{Dust Accretion}      \label{sss:d_accretion}

During galactic evolution, dust grains in the ISM can grow in size due to accretion by metal gas particles on the surface of these grains. This process, occurring mostly in the coldest and densest regions of the ISM, i.e. molecular clouds, has the power to increase the global amount of interstellar dust. For this reason, dust accretion is a fundamental ingredient in dust chemical evolution, as pointed out by many studies (e.g. \citealt{Dwek98}; \citealt{Asano13}; \citealt{Mancini15}).\\
The dust accretion term $\dot{G}_{i,dust,accr}(t)$ can be expressed by means of a typical timescale for accretion $\tau_{accr}$ \citep{Calura08}:
\begin{equation}
\dot{G}_{i,dust,accr}(t)=\frac{G_{i,dust}(t)}{\tau_{i,accr}}.
\label{e:dust_accr}
\end{equation}
\cite{Hirashita00} expressed the dust accretion timescale for the $i$-th element as follows:
\begin{equation}
\tau_{i,accr}=\frac{\tau_g}{X_{cl} (1 - f_i)}.
\end{equation}
In the latter equation, $f_i=G_{i,dust}(t)/G_i(t)$  is the DGR for the element $i$ at the time $t$, while $X_{cl}$ represents the mass fraction of molecular clouds in the ISM. $\tau_g$ is the characteristic dust growth timescale. In our models we adopt the relation given by \cite{Asano13}, who expressed the dust growth timescale $\tau_g$ in a molecular cloud as a function of the metallicity $Z$ as:
\begin{equation}
\tau_g=2.0\cdot 10^7yr \cdot \bigg( \frac{Z}{0.02} \bigg)^{-1},
\end{equation}
assuming $50$ $K$ for the typical molecular cloud temperature, $100$ $cm^{-3}$ for the cloud ambient density and an average value of $0.1$ $\mu m$ for the grain size.

\subsubsection{Dust Destruction}    \label{sss:d_destruction}

Dust grains are not only accreted, but experience also destruction in the ISM. The most efficient process among those able to cycle dust back into the gas phase is the destruction by SN shocks.\\
Similarly to Equation \eqref{e:dust_accr} for dust accretion, $\dot{G}_{i,dust,destr}(t)$ is expressed in terms of the grain destruction timescale $\tau_{destr}$ \citep{Calura08}:
\begin{equation}
\dot{G}_{i,dust,destr}(t)=\frac{G_{i,dust}(t)}{\tau_{destr}}.
\label{e:dust_destr}
\end{equation}
This timescale is assumed to be the same for all the elements depleted in dust and has the following form:
\begin{equation}
\tau_{destr}=\frac{M_{ISM}}{(\epsilon \cdot M_{swept}) SN_{rate}},
\end{equation}
where $M_{swept}$ is the ISM mass swept by a SN shock and $\epsilon$ is the efficiency of grain destruction in the ISM. For the last two parameters, in our model the \cite{Asano13} prescriptions are adopted. They suggest an efficiency $\epsilon=0.1$ and predict for the swept mass:
\begin{equation}
M_{swept}=1535 \cdot \big(Z/Z_\odot+0.039\big)^{-0.289} M_\odot,
\end{equation}
assuming $1cm^{-3}$ for the environment. 


\section{Results}       \label{s:results}

In this Section we attempt to identify the type of the GRB host galaxies present in our sample by comparing the results given by the chemical evolution models with the abundances measured in the GRB hosts. Following the idea developed in the previous works of \citet{Calura09} and \citet{Grieco14}, the procedure consists in comparing model predictions for $[X/Fe]$ vs. $[Fe/H]$ relations with the same relations observed in the GRB afterglow spectra for several chemical elements. In this way, it is possible to reconstruct the star formation history and therefore the morphological type of the galaxy hosting the GRBs.

\subsection{Model Specifications}       \label{ss:models}

\begin{figure}
\centering
	\includegraphics[width=0.54\columnwidth]{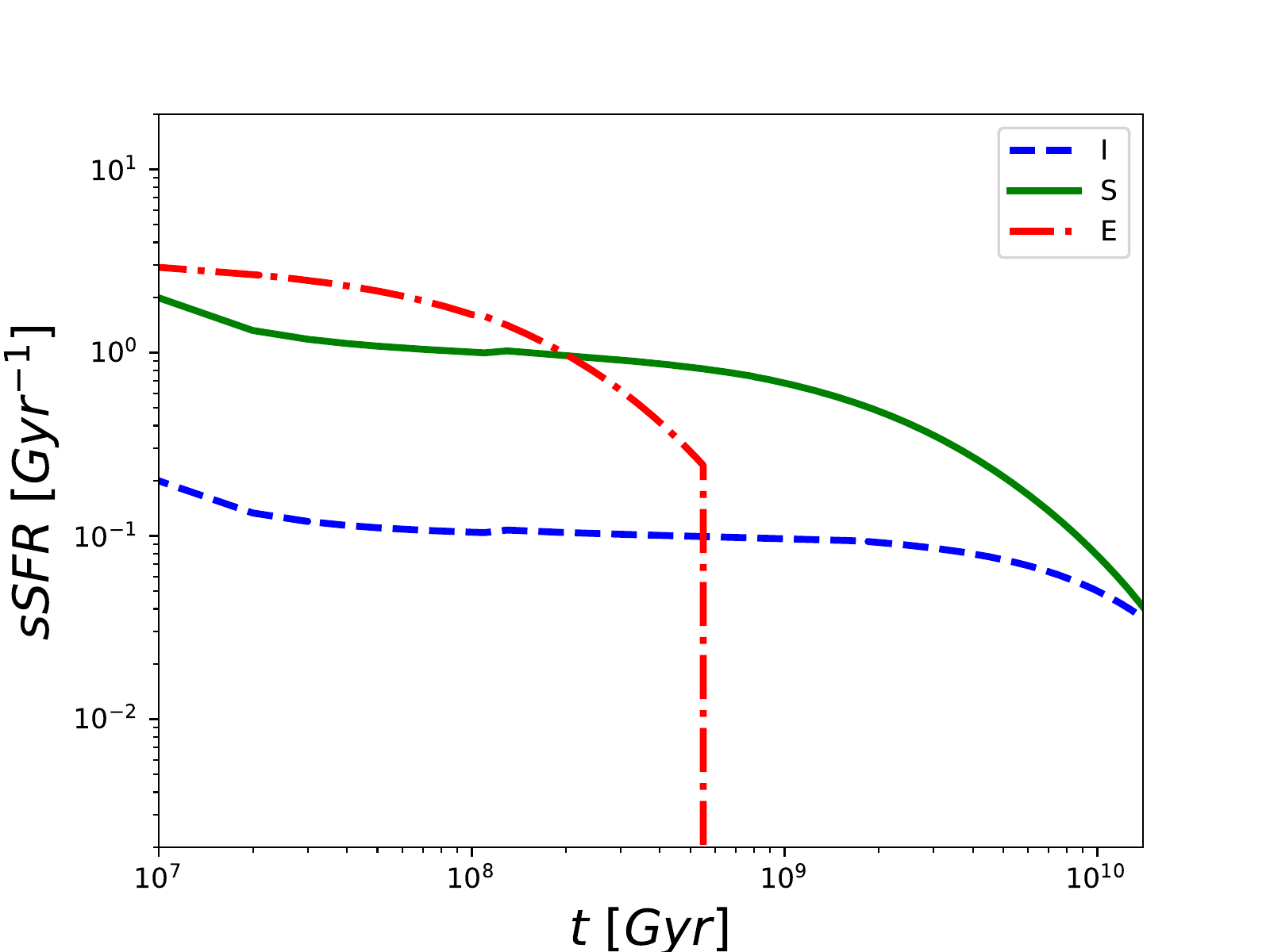}
    \caption{Predicted sSFR as a function of time for the reference models of galaxies of different morphological types. S stands for spiral, I for irregular and E for spheroid (locally elliptical).}
    \label{f:SFR}
\end{figure}

The first step in the identification of the GRB hosts is to build chemical evolution models, following the evolution of the chemical abundances of several elements in the gas and dust. 
Models are aimed at reproducing the observed main features of local galaxies: in particular, we build reference models for typical galaxies of each morphological type. In this way, we are confident of adopting realistic models for each galaxy morphology and to be able to follow the galaxy evolution of each type since the beginning and until the present time.\\
We develop models for i) a typical spheroid, ii) a typical spiral Milky Way-like and iii) a typical irregular. The main differences among the models for different galaxy types are the assumed star formation efficiency (SFE, Equation \ref{e:SFR}) and  the timescale for gas infall, which determine the history of star formation, the main driver of galaxy evolution. In Table \ref{t:models1} the basic assumptions (infall mass, infall timescale, SFE, galactic wind parameter and dust prescriptions) are shown for the reference models. The IMF adopted in the reference models is that of \citet{Salpeter55} for all galaxy models.
In the last column of Table \ref{t:models1}, the choice of the condensation efficiencies $\delta_i^{CC}$ for CC-SNe (see \ref{ss:d_prod}) is specified: $\delta_{HP}$ stands for the condensation efficiencies with a low density circumstellar environment ($n_H = 0.1$ $cm^{-3}$ ), that leads to higher $\delta_i$ values, whereas $\delta_{MP}$ stands for the prescription with $n_H = 1cm^{-3}$ density, that implies lower dust production by massive stars.\\
In Figure \ref{f:SFR} we show the predicted sSFR (specific star formation rate, i.e. SFR normalised per unit luminous mass) vs. time for the three reference models. In the case of the spheroidal massive galaxy (locally elliptical), we see a very high initial SFR followed by an abrupt decline due to the occurrence of a galactic wind which devoids the galaxy of the gas residual from star formation. This wind occurs before 1 $Gyr$ since the beginning of star formation, and ensures that the dominant stellar population shows enhanced $[\alpha/Fe]$ ratios as observed \citep{Matteucci94,Pipino04,DeMasi18_1}. In the case of the spiral and irregular galaxies, the SFR behaviour is continuous and the star formation lasts until the present time. We have checked that the present time absolute SFRs for a spiral and irregular galaxy reproduce the values found for the Milky Way disk ($SFR_{MW}= 1.9 \pm {0.4} M_{\odot}yr^{-1}$, \citealt{Chomiuk11}) and SMC ($SFR_{SMC}= 0.053^{+0.03}_{-0.02} M_{\odot}yr^{-1}$, \citealt{Rubele15}), respectively. These histories of star formation had been already tested in previous papers \citep{Grieco12,Gioannini17b} and ensure us that we can reasonably reproduce a galaxy of a given morphological type. However, since real galaxies show a distinctive spread in their properties we have also considered a range of values for the adopted parameters, as shown in Table \ref{t:models2}. These ranges are chosen in such a way to still reproduce the main chemical properties of local elliptical (\citealt{Pipino04,DeMasi18_1}), spiral (\citealt{Chiappini01}; \citealt{Cescutti07}) and irregular galaxies (e.g. \citealt{Lanfranchi03}, see also \citealt{Calura09b}).\\ 

\begin{table*} 
\centering
\tablenum{2}
\caption{Input parameters for the reference chemical evolution models adopted for galaxies of different morphological type.}
\begin{tabular}{c *{5}c}
\hline
 Model & $M_{inf} [M_\odot]$ & $\tau_{inf} [Gyr]$ & $\nu [Gyr^{-1}]$ & $K_i$ & $\delta^{CC}$\\[0.1cm]
\hline
 Irregular (I) & $5\cdot10^9$ & $10$ & $0.1$ & $0.5$ & $\delta_{HP}$ \\
 Spiral (Sp) & $5\cdot10^{10}$ & $7$ & $1$ & $0.2$ & $\delta_{MP}$ \\
 Spheroid (E) & $10^{11}$ & $0.3$ & $15$ & $10$ & $\delta_{MP}$ \\
\hline
\end{tabular}
\label{t:models1}
\end{table*}

\begin{table*} 
\centering
\tablenum{3}
\caption{Ranges of input parameters explored for the chemical evolution models for galaxies of different morphological types.}
\begin{tabular}{c *{5}c}
\hline
 Models & ${M_{inf} [M_\odot]}$ & ${\tau_{inf} [Gyr]}$ & ${\nu [Gyr^{-1}]}$ & ${K_i}$ & ${\delta^{CC}}$\\[0.1cm]
\hline
 Irregulars & ${5\cdot10^8<M<5\cdot10^9}$ & ${9<\tau<11}$ & ${0.01<\nu<0.2}$ & ${0.5<K<1}$ & ${\delta_{HP}}$, ${\delta_{MP}}$ \\
 Spiral disks & ${10^{10}<M<10^{11}}$ & ${4<\tau<8}$ & ${1<\nu<3}$ & ${0.1<K<0.5}$ & ${\delta_{HP}}$, ${\delta_{MP}}$ \\
 Spheroids & ${10^{11}<M<10^{12}}$ & ${0.2<\tau<0.5}$ & ${10<\nu<25}$ & ${10<K<20}$ & ${\delta_{HP}}$, ${\delta_{MP}}$ \\
\hline
\end{tabular}
\label{t:models2}
\end{table*}

\begin{figure*}
\centering
\includegraphics[width=.58\textwidth]{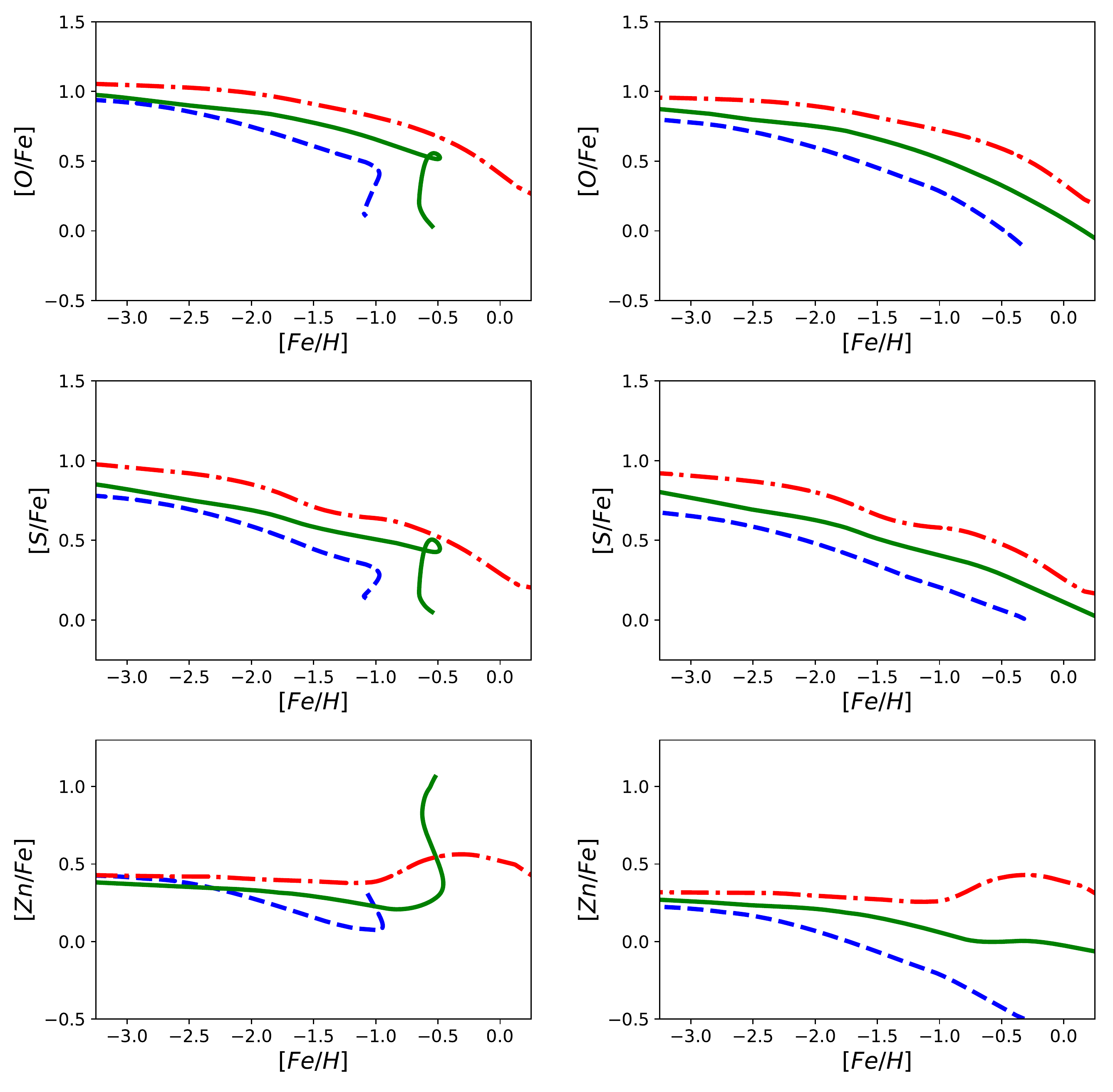}
\centering
\caption{Example of $[X/Fe]$ vs. $[Fe/H]$ ratios behaviour for the chemical evolution models adopted in this work. The blue dashed (I), green solid (Sp) and red dash-dotted lines (E) are the predictions computed by means of reference models for an irregular, a spiral and an spheroidal galaxy. The left panels show the models considering dust, whereas the right panels the models without 
dust.} 
\label{f:GRB_nodust}
\end{figure*}

\noindent Before starting with the identification, some other model features need to be mentioned. First, in this work we assume for $Ni$ (an element not considered in \citealt{Piovan11}) the same condensation efficiencies as for $Fe$. This solution is reasonable, due to the very similar condensation temperatures of the two elements \citep{Taylor01} and the fact that $Ni$ belongs to the so-called $Fe$-peak group. In our work we consider $C$, $O$, $Mg$, $Si$, $S$, $Fe$ and $Ni$ as condensed in dust
, whereas for $Zn$ and $N$ we assume no dust depletion, since it is known that these two elements are volatile. 
In Figure \ref{f:GRB_nodust} we show what happens by considering dust in our reference models: in particular, we show the behaviour of some abundance ratios as functions of $[Fe/H]$ in the presence and in absence of dust. As one can see, the effects of condensation into dust are quite important. In fact, dust production by stars increases the $[X/Fe]$ ratios for the elements  of Figure \ref{f:GRB_nodust} at very low metallicities. Moving toward higher metallicities, the abundance patterns are completely changed because of dust accretion, the main contributor to dust mass at late times \citep{Gioannini17b,Ginolfi18}. The effects caused by dust accretion are not visible for the spheroid model: this happens because of the galactic wind, which completely devoids the galaxy of its gas and dust content.\\
Regarding $N$, for this paper we ran all the models twice. First we adopted \citet{Nomoto13} yields for $N$, which do not consider primary production from massive stars, and then we used \citet{Matteucci86} prescription, considering instead a fixed amount of primary $N$ produced by massive stars, irrespective of the stellar metallicity. This assumption is ad hoc but it reproduces the $[N/Fe]$ ratios in the solar vicinity (in particular the observed plateau at low metallicities in MW halo stars, \citealt{Matteucci86}), as well as in low metallicity QSO-DLAs (\citealt{Pettini02}, \citeyear{Pettini08}) and in low metallicity star forming galaxies (e.g. \citealt{Berg12,James15}). This behaviour is in contrast with what is expected from standard nucleosynthesis models, such as the \citet{Nomoto13} ones, predicting only secondary $N$ to be produced by massive stars ($m\gtrsim 10M_\odot$). With secondary production only, in fact, a quadratic evolution with metallicity is obtained (see \citealt{Maiolino19}). On the other hand, primary $N$ production has been predicted in rotating very low metallicity massive stars (e.g. \citealt{Meynet02,Frischknecht16}) but unfortunately these models do not predict primary $N$ for higher metallicity rotating stars, as instead is required by observations. For this reasons, we adopt only the \citet{Nomoto13} and the \citet{Matteucci86} scenarios in the hope of better understanding the nature of nitrogen.

\subsection{Host Identification}    \label{ss:host_id}

In this Subsection, we see what happens by adopting a \citet{Salpeter55} IMF for all the models. However, as for the long-dated debate on the universality of the IMF (\citealt{Kroupa02}; \citealt{Ferreras16}), in the next Subsection we will also see what are the effects of adopting  different IMFs in different morphological types.\\
Returning to GRB hosts identification, we adopt a statistical test, already used in the works of \citeauthor{Dessauges04} (\citeyear{Dessauges04}, \citeyear{Dessauges07}). This test consists in determining the minimal distance between the data points and the curve of the model representing the $[X/Fe]$ vs. $[Fe/H]$ relation. In particular, we derived this minimal distance by looking at the distance $d_X$ for which the ratio $d_X/\sigma_X$, where $\sigma_X$ is the error for the abundance data, is minimal. After that, we computed the weighted mean for all the abundance diagrams considered in each system. From the comparison of these means, we obtained the best model representing the GRB host. This procedure also gives the opportunity to approximately estimate the age of the host galaxy (namely the time passed since the very first episode of star formation). Each point of minimal distance inferred for the best model corresponds to a time $t_X$. By weighting these times on the reciprocals of the ratios $d_X/\sigma_X$, we derived the age of the host. Upper and lower limits are not taken into account in this procedure.\\
We consider in our statistical test all elements except $Ni$ for which the theoretical stellar yields are very uncertain and unable to reproduce the solar vicinity data (\citealt{Koba06,Romano10}, adopting the same yields of $Ni$ as used here). 
Unfortunately, even newer sets of yields (e.g. \citealt{Limongi18}) do not allow to reproduce $Ni$ evolution, as shown by \citet{Prantzos18} for the solar neighbourhood.

\subsubsection{GRB 050730}      \label{sss:050730}

\begin{figure*}
\centering
\includegraphics[width=.81\textwidth]{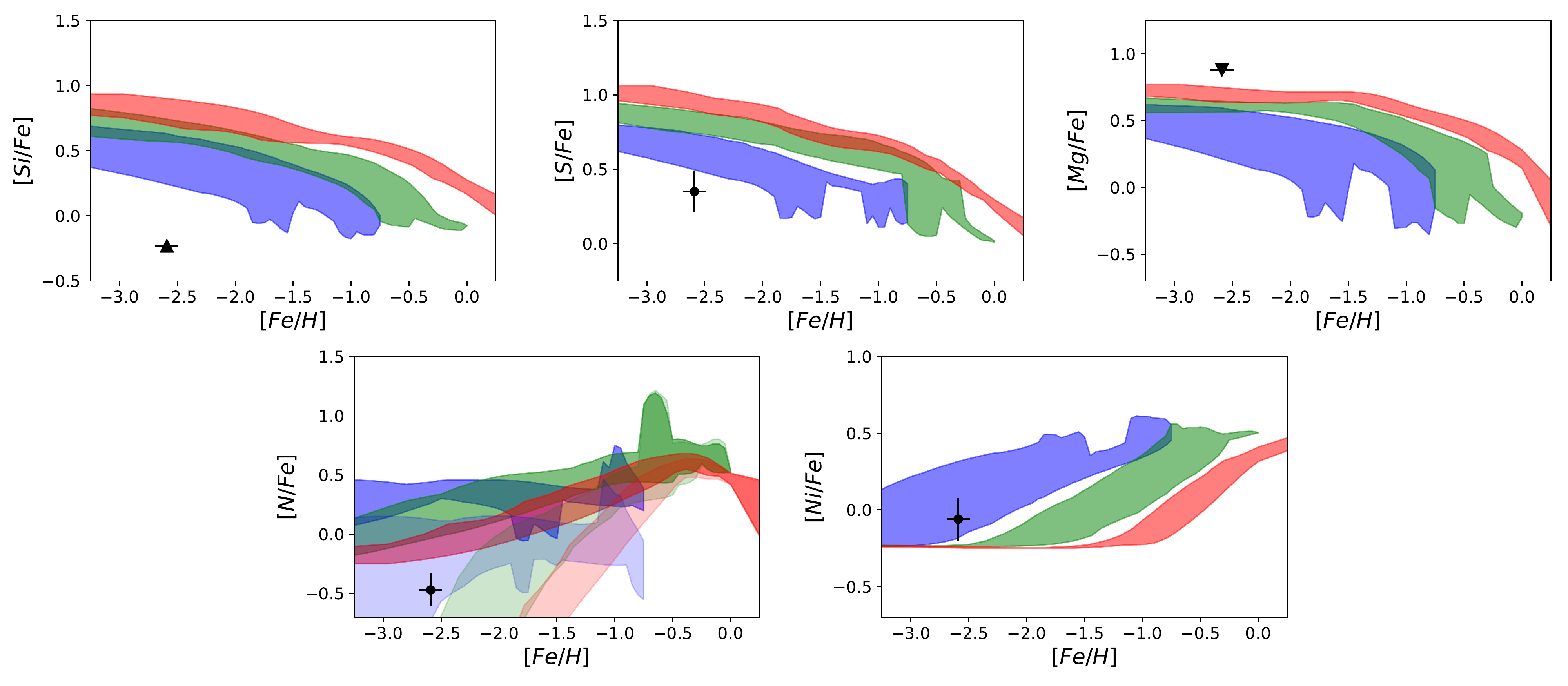}
\centering
\caption{Observed $[X/Fe]$ vs. $[Fe/H]$ ratios for the GRB 050730 host galaxy provided by \citet{Prochaska07}. Data are black symbols with error bars; up and down triangles indicate lower and upper limits. The blue, green and red shaded areas are the predictions computed by means of models for irregular, spiral and spheroidal galaxies, respectively. In the lower left panel we show the results considering both primary $N$ production from massive stars (\citealt{Matteucci86}, darker shaded areas) and only secondary $N$ production (\citealt{Nomoto13}, lighter shaded areas).}
\label{f:050730}
\end{figure*}

\begin{figure*}
\centering
\includegraphics[width=.81\textwidth]{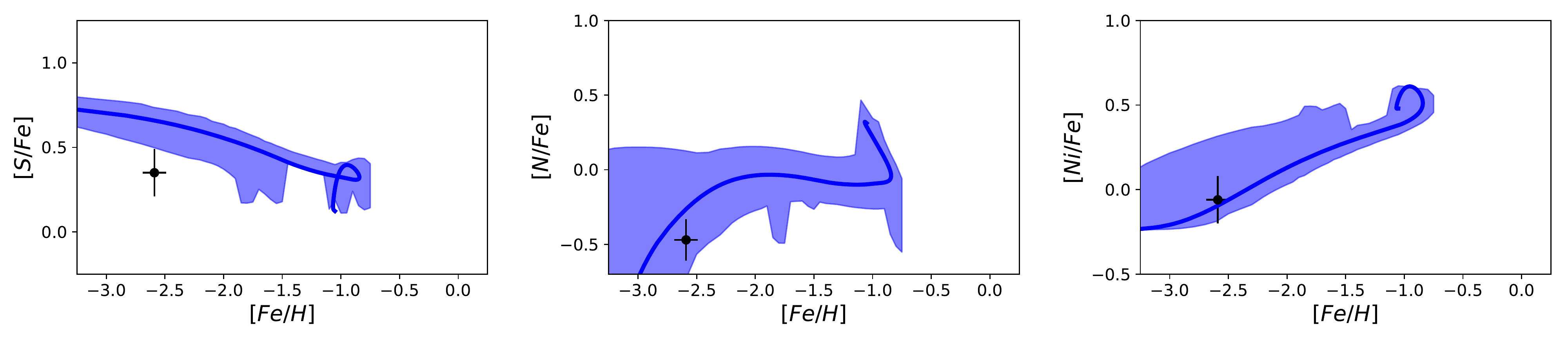}
\caption{Observed $[X/Fe]$ vs. $[Fe/H]$ ratios for the GRB 050730 host galaxy provided by \citet{Prochaska07}. Data are black symbols with error bars. The blue shaded area and the blue line are the predictions computed by means of all irregular galaxy models and the best model for the GRB host, respectively. In the central panel we show the results considering only secondary $N$ production from massive stars \citep{Nomoto13}.} 
\label{f:050730_1}
\end{figure*}

In Figure \ref{f:050730}, we show the results for different morphological type models, with the parameters presented in Table \ref{t:models2}. In this Figure, we also present the abundance ratios measured by \citet{Prochaska07} for the GRB 050730 afterglow.\\
Looking at the Figure, the analysis suggests that the host galaxy has a SFH typical of an irregular galaxy. $[S/Fe]$ and $[Ni/Fe]$ are particularly suggesting this hypothesis (although $Ni$ has the problems described before), corroborated by the compatibility with $Si$ and $Mg$ lower and upper limits, respectively. Concerning $[N/Fe]$, the observed ratio is in agreement with irregular models considering only secondary production by massive stars (\citealt{Nomoto13}, lighter shaded areas). Yields with primary production by \citet{Matteucci86}, instead, give for all the adopted models too high values relative to the observations.\\
Figure \ref{f:050730_1} shows the best model for this GRB host. We find that the best model to describe the observed ratios is the one for an irregular galaxy with moderate SFE ($0.1Gyr^{-1}$). We also find that the lowering the dust production by massive stars (relative to the irregular reference model, i.e. $\delta_{MP}$ instead of $\delta_{HP}$) gives a slightly better data-model agreement.\\
In this way, adopting the age determination procedure described in Subsection \ref{ss:host_id}, we find for the host an age of $\sim 0.2$ $Gyr$.

\subsubsection{GRB 050820}      \label{sss:050820}


\begin{figure*}
\centering
\includegraphics[width=.81\textwidth]{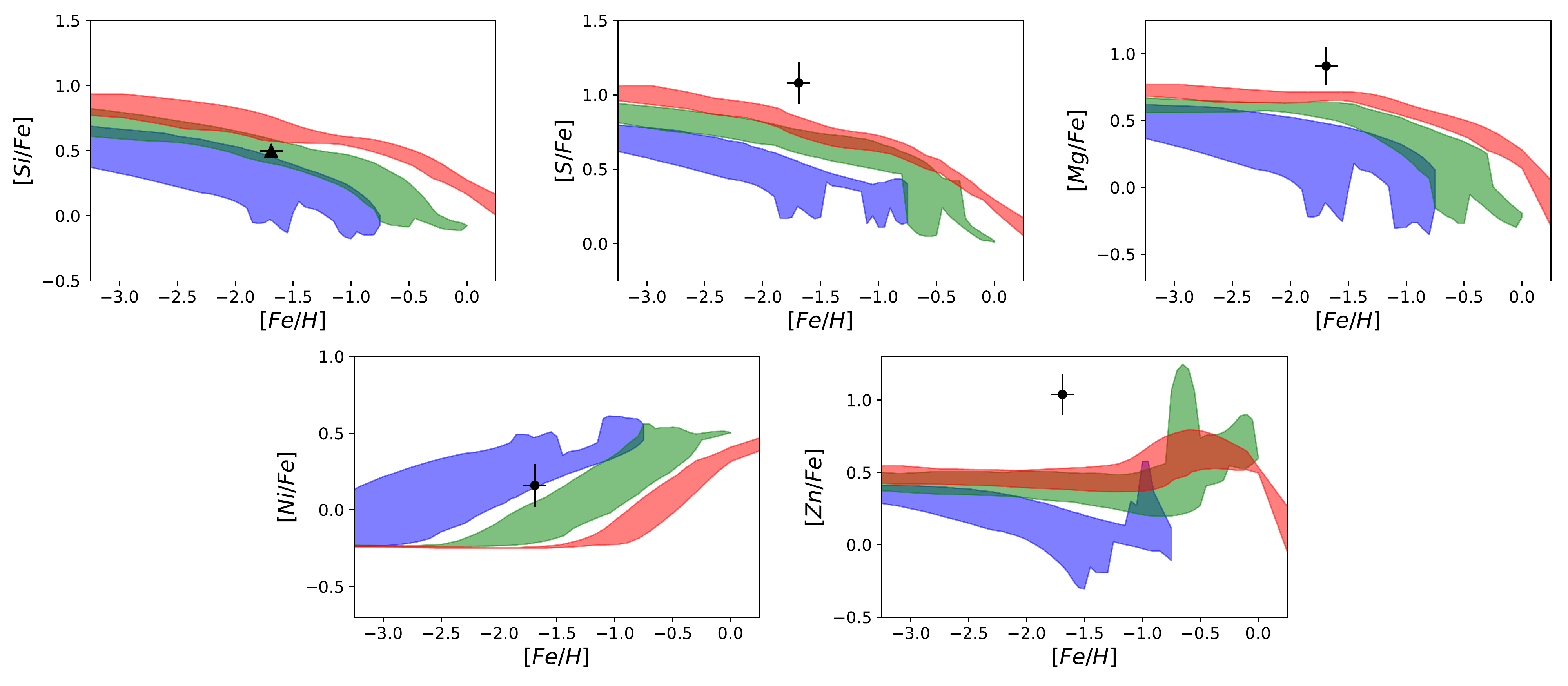}
\centering
\caption{Observed $[X/Fe]$ vs. $[Fe/H]$ ratios for the GRB 050820 host galaxy provided by \citet{Prochaska07}. Data are black symbols with error bars; up triangles indicate lower limits. The blue, green and red shaded areas are the predictions computed by means of models for irregular, spiral and spheroidal galaxies, respectively.}
\label{f:050820}
\end{figure*}

\begin{figure*}
\centering
\includegraphics[width=.81\textwidth]{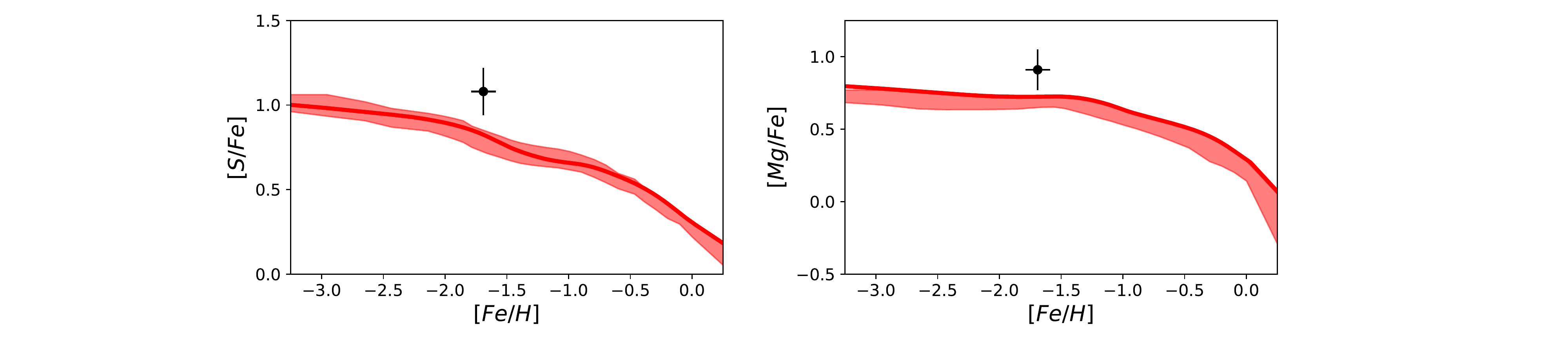}
\centering
\caption{Observed $[X/Fe]$ vs. $[Fe/H]$ ratios for the GRB 050820 host galaxy provided by \citet{Prochaska07}. Data are black symbols with error bars. The red shaded area and the red line are the predictions computed by means of all spheroidal galaxy models and the best model for the GRB host, respectively.} 
\label{f:050820_1}
\end{figure*}

In Figure \ref{f:050820}, we see that the three upper panels for $[\alpha/Fe]$ ratios seem to indicate a star forming spheroid for this host. At the same time, the panel showing $[Ni/Fe]$ vs. $[Fe/H]$ indicates late-type (irregular or spiral) galaxy models, but we do not consider this element as a discriminant because of its uncertain yields. For $[Zn/Fe]$, instead, we note that the observed value is much higher than all the model predictions. We will see however that the observed overabundance of this ratio relative to what is predicted by the models is common, in particular for the host identified as spheroids.
We note that $Zn$ seems to behave like an $\alpha$ element: in this and in other systems $[Zn/Fe]$ is enhanced when other $\alpha$-elements (i.e. $Si$, $S$, $Zn$) are enhanced. 
This similarity between $\alpha$ elements and $Zn$ is an important observational fact, since it casts doubts on the assumption that $Zn$ traces $Fe$, as suggested in the literature \citep{DeCia16}. As a matter of fact, the large difference between the data and the predictions ($\sim 0.5 dex$) can be hardly explained in terms of dust depletion. In fact, even artificially increasing $Fe$ dust yields, the observed values are not reached. At the same time, $Fe$ dust growth seems not to be the solution: we need too small accretion timescales to be physically realistic.\\ 
In Figure \ref{f:050820_1}, we show the results of the model with very high SFE and infall mass ($\nu=25Gyr^{-1}$, $M_{inf}=10^{12}M_\odot$), which results to be the best from the statistical test. As expected from the $\lq\lq$time-delay model", $[\alpha/Fe]$ ratios tend to rise with the SFE (and so the SFR).\\
Concluding, we identify this host galaxy as a high mass and strong star forming spheroidal galaxy. For what concerns the age of this host, we find a very young one of $\sim 15$ $Myr$. 
\begin{figure}
\centering
\includegraphics[width=0.5\columnwidth]{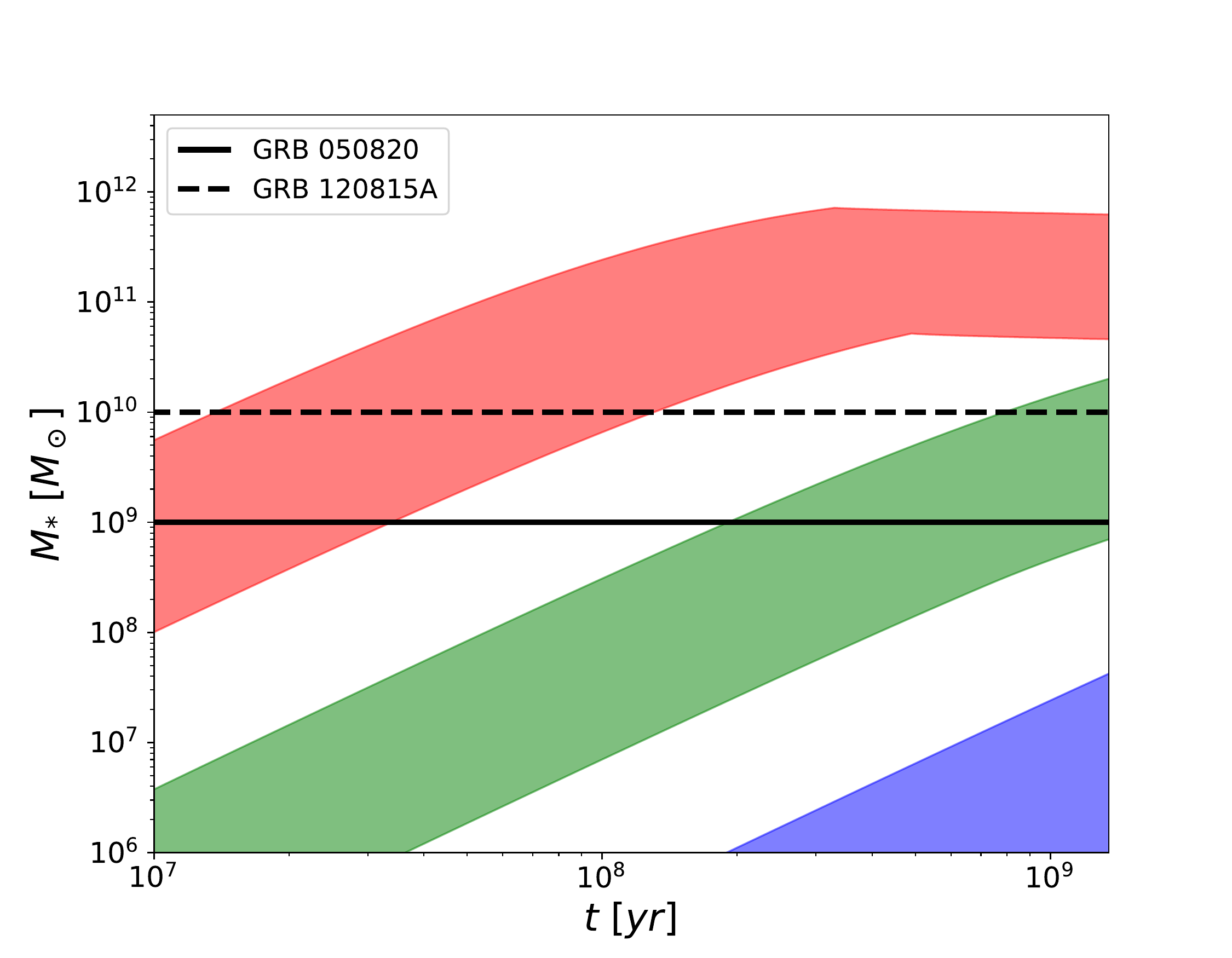}
\caption{Stellar mass evolution for GRB models adopted in this work. The blue, green and red shaded areas are the predictions computed by means of models for irregular, spiral and spheroidal galaxies, respectively. The black solid and dashed lines are the mass estimates for GRB 050820 and GRB 120815A hosts, respectively.}
\label{f:mass_host}
\end{figure}
This very young galactic age allows us to predict for this host a stellar mass that is roughly consistent with the observationally inferred one (see Figure \ref{f:mass_host}).

\begin{figure*}
\centering
\includegraphics[width=.81\textwidth]{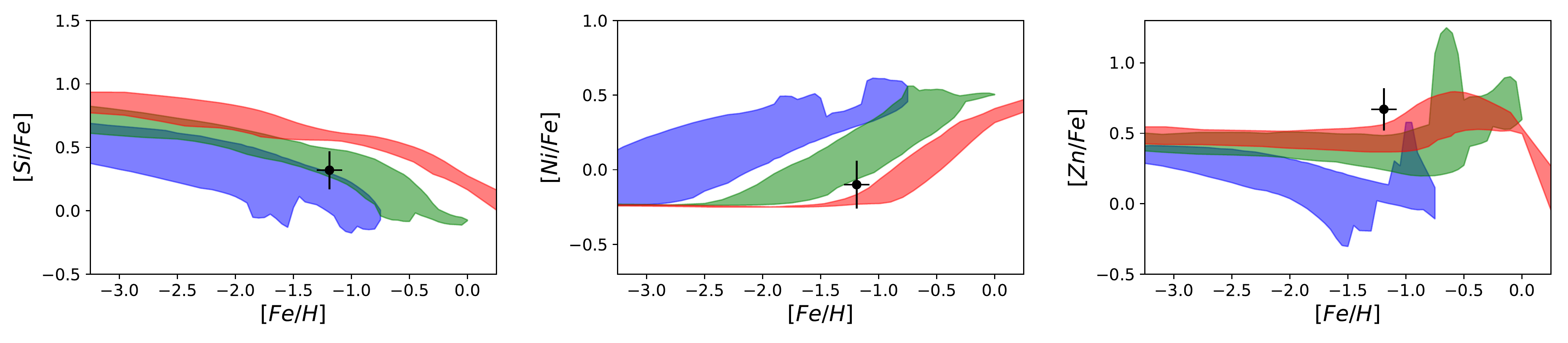}
\centering
\caption{Observed $[X/Fe]$ vs. $[Fe/H]$ ratios for the GRB 081008 host galaxy provided by \citet{DElia11}. Data are black symbols with error bars. The blue, green and red shaded areas are the predictions computed by means of models for irregular, spiral and spheroidal galaxies, respectively.}
\label{f:081008}
\end{figure*}
\begin{figure*}
\centering
\includegraphics[width=.81\textwidth]{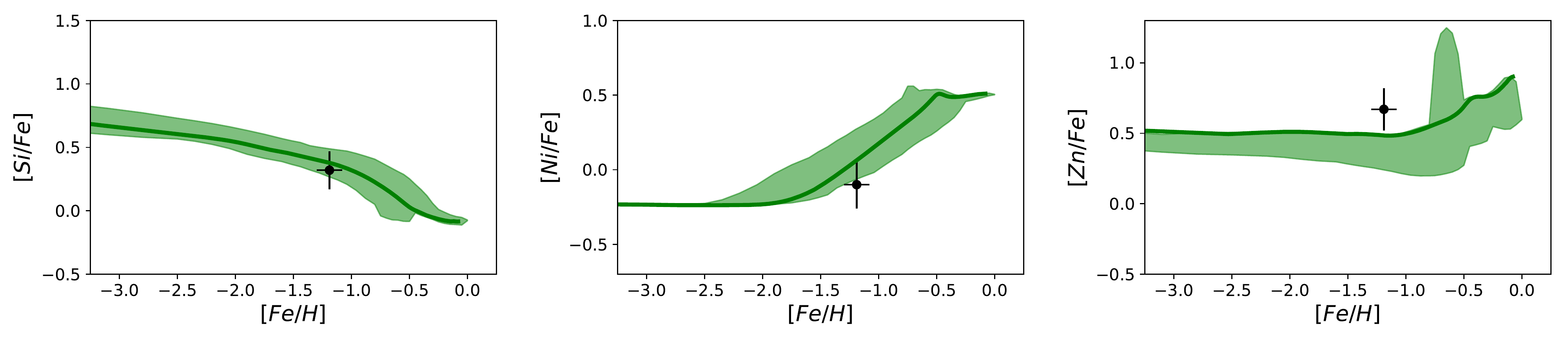}
\caption{Observed $[X/Fe]$ vs. $[Fe/H]$ ratios for the GRB 081008 host galaxy provided by \citet{DElia11}. Data are black symbols with error bars. The green shaded area and the green line are the predictions computed by means of all spiral galaxy models and the best model for the GRB host, respectively.} 
\label{f:081008_1}
\end{figure*}

\subsubsection{GRB 081008}      \label{sss:081008}

For this host, as shown in Figure \ref{f:081008}, the observed abundance ratios are quite well fitted by the models for spiral galaxies, in particular looking at $[Si/Fe]$ and $[Ni/Fe]$. The low $[Si/Fe]$ in particular (remember the uncertainties in $Ni$ yields) suggests the idea of a late type galaxy (as the spiral like the MW). 
Spiral models little underpredict the observed $[Zn/Fe]$.\\ 
In Figure \ref{f:081008_1}, the predictions of the best model are shown. The statistical test suggests as the best model the one for a strongly star forming (SFE $= 3 Gyr^{-1}$) spiral disk. This model also assume increased dust production by massive stars relative to the spiral reference model (i.e. $\delta_{HP}$ instead of  $\delta_{MP}$). Some considerations can be drawn for the best model: looking at the $[Zn/Fe]$ vs. $[Fe/H]$ panel, one can explain the observed $[Zn/Fe]$ in terms of dust accretion. In fact, a faster dust accretion could help, alleviating the discrepancy between the model and data.\\
Adopting the strong star forming spiral model as the best one for this host galaxy, we estimate the age of the host at the time of the GRB event to be $\sim0.2$ $Gyr$.

\subsubsection{GRB 090926A}     \label{sss:090926A}

\begin{figure*}
\centering
\includegraphics[width=.81\textwidth]{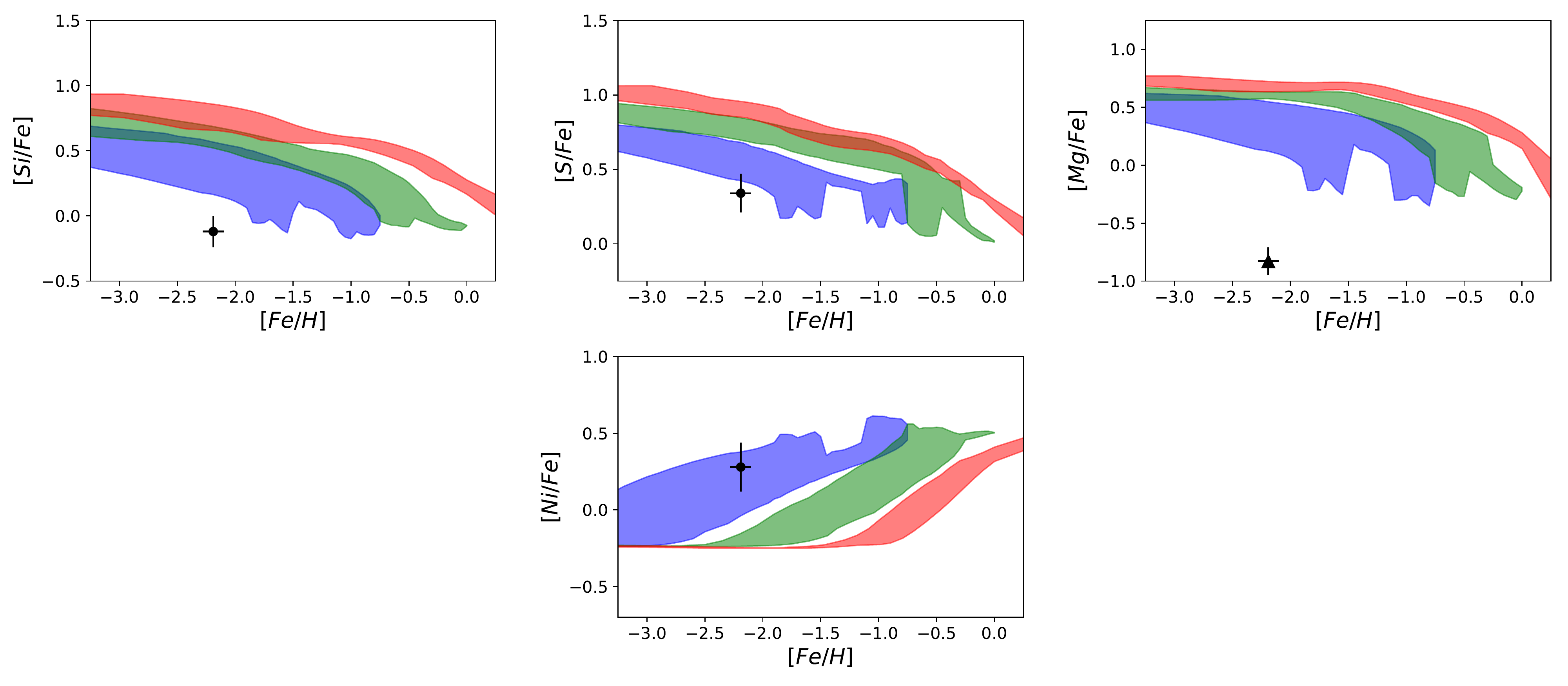}
\centering
\caption{Observed $[X/Fe]$ vs. $[Fe/H]$ ratios for the GRB 090926A host galaxy provided by \citet{DElia10}. Data are black symbols with error bars; up triangles indicate lower limits.  The blue, green and red shaded areas are the predictions computed by means of models for irregular, spiral and spheroidal galaxies, respectively.}
\label{f:090926}
\end{figure*}

\begin{figure*}
\centering
\includegraphics[width=.81\textwidth]{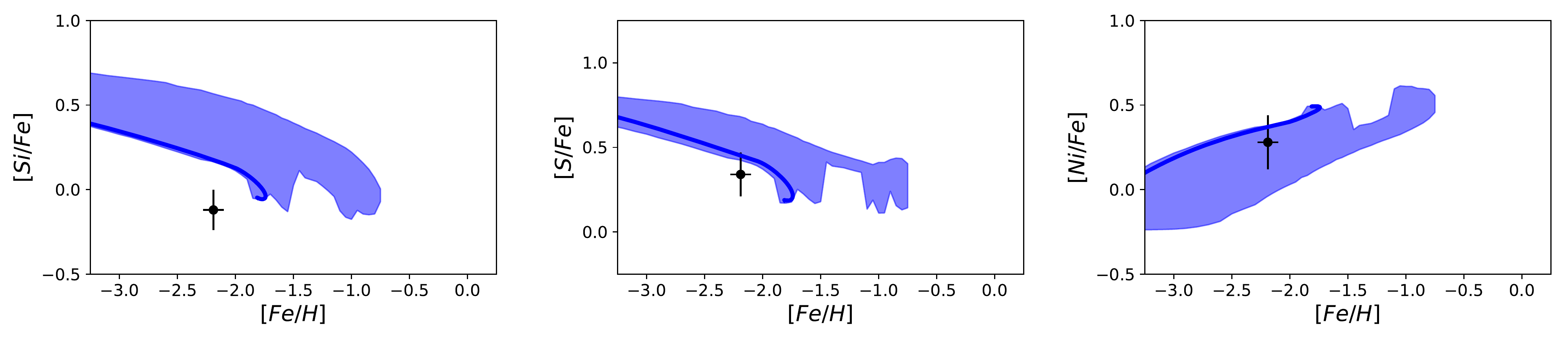}
\caption{Observed $[X/Fe]$ vs. $[Fe/H]$ ratios for the GRB 090926A host galaxy provided by \citet{DElia10}. Data are black symbols with error bars. The blue shaded area and the blue line are the predictions computed by means of all irregular galaxy models and the best model for the GRB host, respectively.} 
\label{f:090926_1}
\end{figure*}

In Figure \ref{f:090926}, the three upper panels showing $[\alpha/Fe]$ vs. $[Fe/H]$ indicate very low abundance values for $\alpha$-elements (although $Mg$ is a lower limit). This fact, coupled with a low $[Fe/H]$, suggests an irregular galaxy. Also $[Ni/Fe]$ plot in the lower right panel is in good agreement with the irregular models range of values. We do not show in Figure \ref{f:090926} the comparison between the predicted and observed $[N/Fe]$. As anticipated in Table \ref{t:observation}, the abundance determination for $N$ in this object was affected by severe problems. In fact, in \citet{DElia10} it was clearly stated that the low measured $N$ abundance is very probably a lower limit. For this reason, we exclude this element from our analysis.\\
Among all the models considered in Table \ref{t:models2}, we find that the best one has very low infall mass ($5\cdot 10^8M_\odot$) and SFE ($0.01 Gyr^{-1}$). The predictions of this model are shown in Figure \ref{f:090926_1}.
It is evident from the left and central panels that we have better agreeement with data by lowering as much as possible the SFE and the mass, beacause it lowers the $[\alpha/Fe]$ ratios. Still looking at Figure \ref{f:090926_1}, we see that $[S/Fe]$ observed ratio is more in agreement with the best model than $[Si/Fe]$. Possible explanations of this difference can be found by looking at the dust amount of $Si$ and $S$ relative to $Fe$, or for the effects of a differential wind (different mass loading factor for different elements).\\
When adopting the above low mass and SFE irregular galaxy model, we find that this host is $\sim 1.3$ $Gyr$ old.  This makes the host of GRB 090926A the oldest in terms of galactic age among those studied in this work.
 

\subsubsection{GRB 120327A}     \label{sss:120327A}

\begin{figure*}
\centering
\includegraphics[width=.81\textwidth]{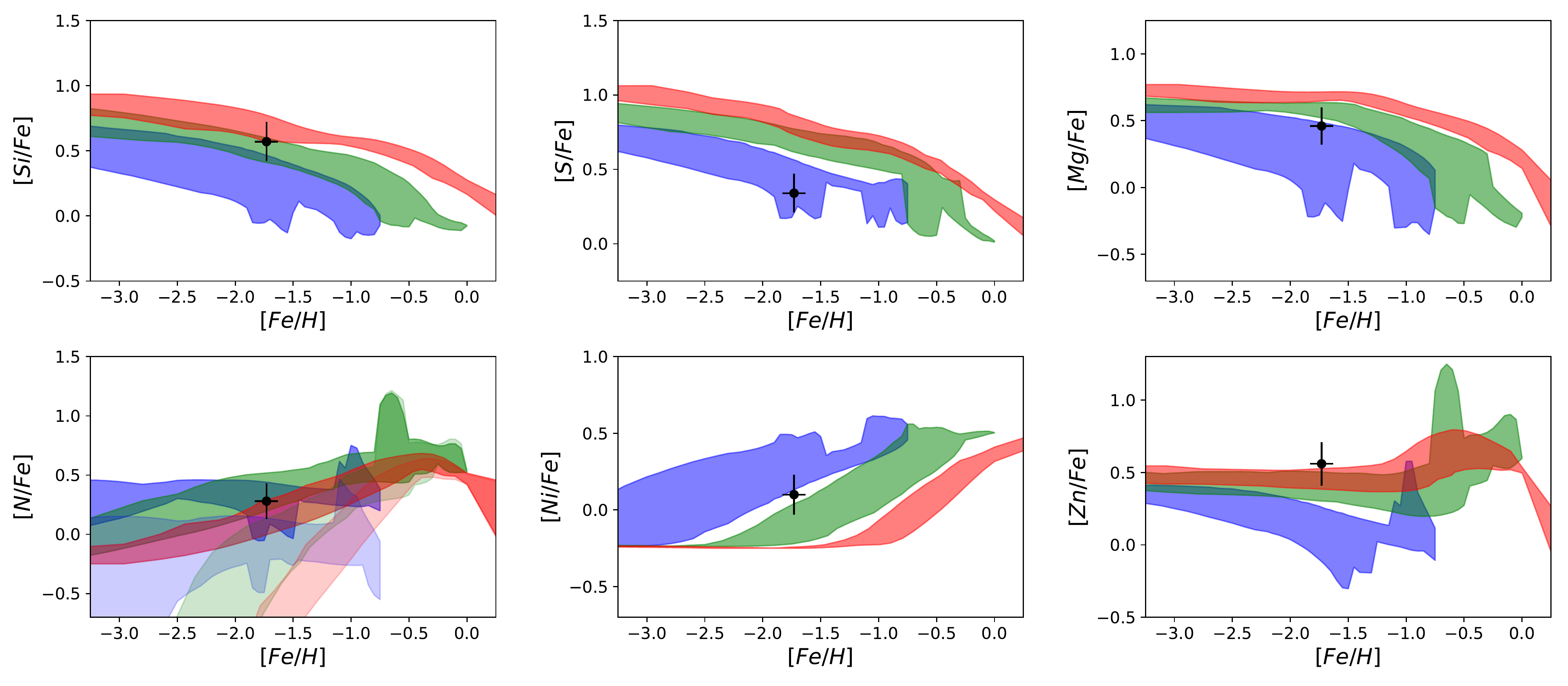}
\centering
\caption{Observed $[X/Fe]$ vs. $[Fe/H]$ ratios for the GRB 120327A host galaxy provided by \citet{DElia14}. Data are black symbols with error bars. The blue, green and red shaded areas are the predictions computed by means of models for irregular, spiral and spheroidal galaxies, respectively. In the lower left panel we show the results considering both primary $N$ production from massive stars (\citealt{Matteucci86}, darker shaded areas) and only secondary $N$ production (\citealt{Nomoto13}, lighter shaded areas).} 
\label{f:120327}
\end{figure*}
\begin{figure*}
\centering
\includegraphics[width=.81\textwidth]{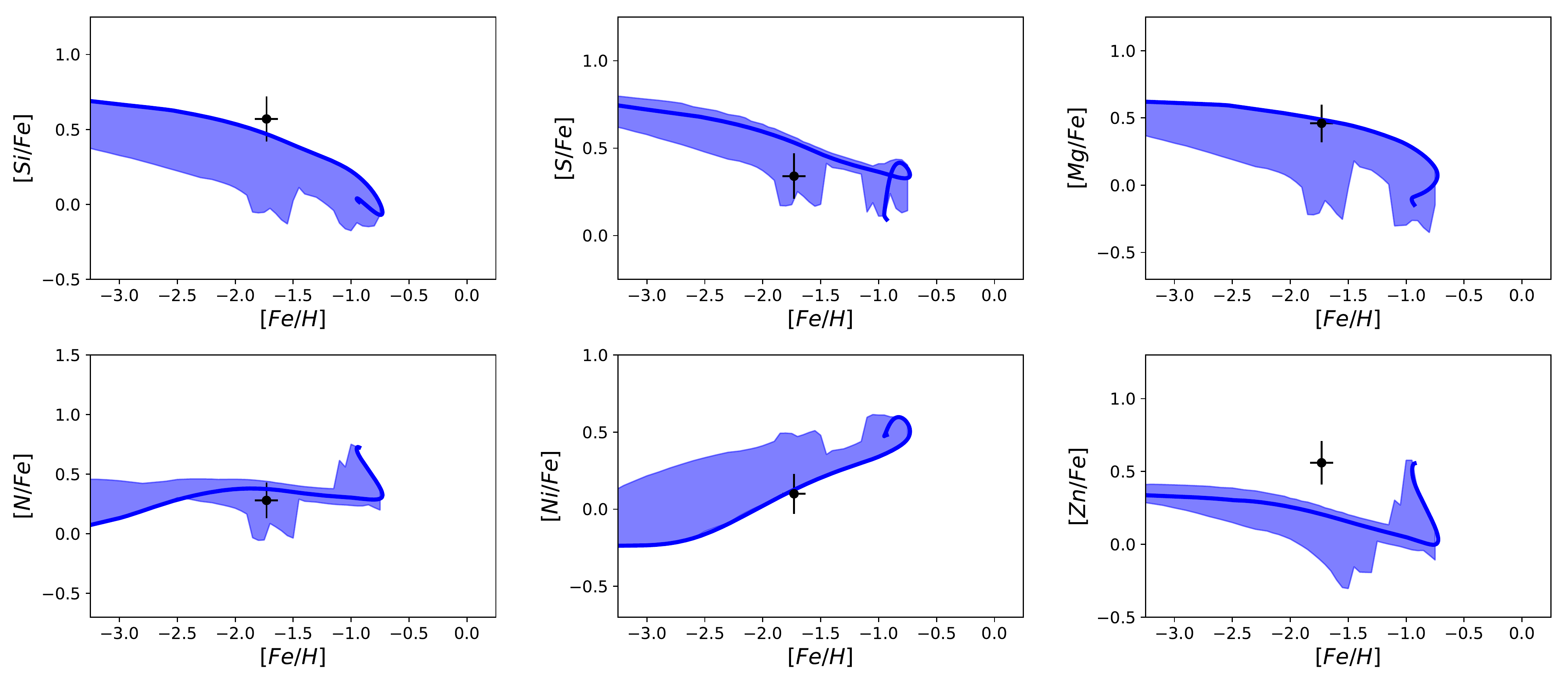}
\caption{Observed $[X/Fe]$ vs. $[Fe/H]$ ratios for the GRB 120327A host galaxy provided by \citet{DElia14}. Data are black symbols with error bars. The blue shaded area and the blue line are the predictions computed by means of all irregular galaxy models and the best model for the GRB host, respectively. In the lower left panel we show the results considering primary $N$ production from massive stars \citep{Matteucci86}.}
\label{f:120327_1}
\end{figure*}
\begin{figure*}
\centering
\includegraphics[width=.81\textwidth]{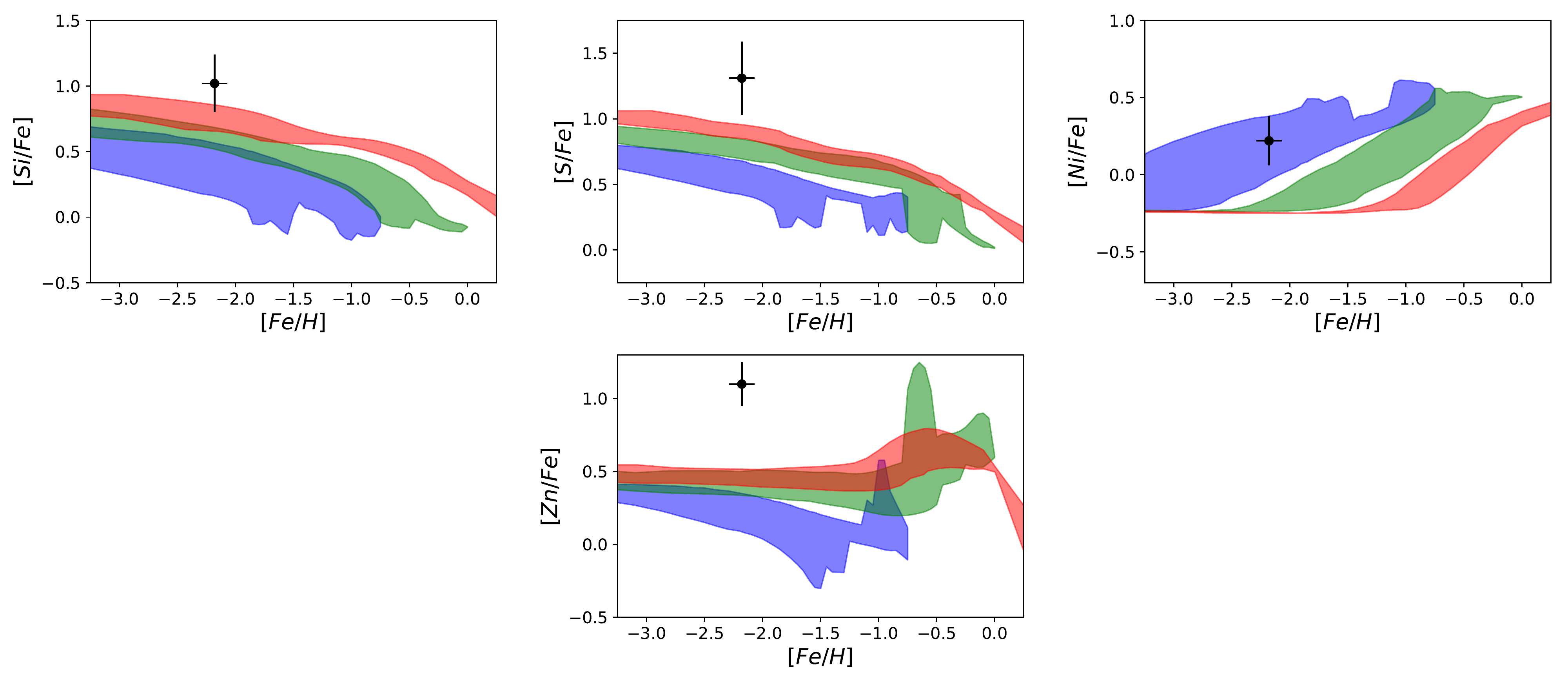}
\centering
\caption{Observed $[X/Fe]$ vs. $[Fe/H]$ ratios for the GRB 120815A host galaxy provided by \citet{Kruhler13}. Data are black symbols with error bars. The blue, green and red shaded areas are the predictions computed by means of models for irregular, spiral and spheroidal galaxies, respectively.}
\label{f:120815}
\end{figure*}
\begin{figure*}
\centering
\includegraphics[width=.81\textwidth]{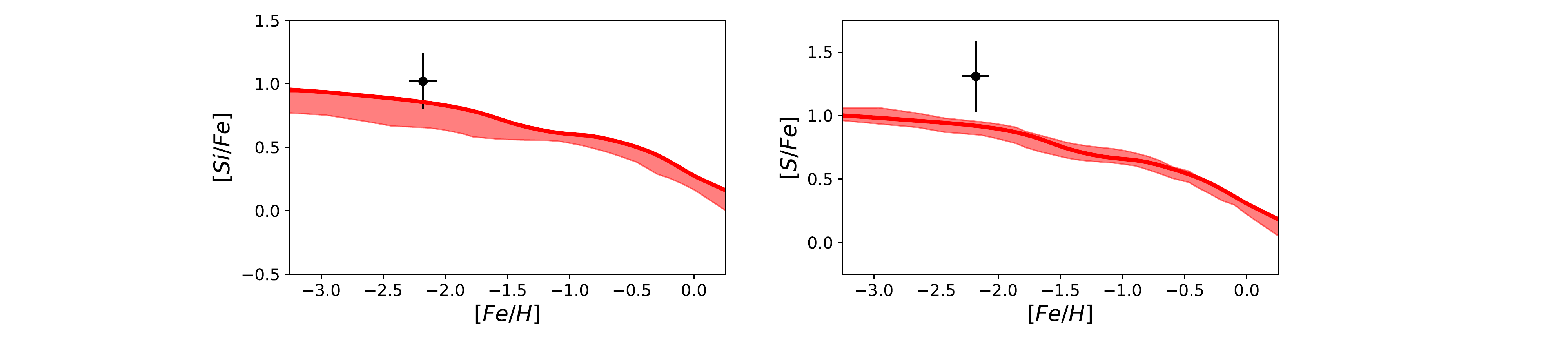}
\caption{Observed $[X/Fe]$ vs. $[Fe/H]$ ratios for the GRB 120815A host galaxy provided by \citet{Kruhler13}. Data are black symbols with error bars. The red shaded area and the red line are the predictions computed by means of all  spheroidal galaxy models and the best model for the GRB host, respectively.}
\label{f:120815_1}
\end{figure*}

The comparison of the patterns for the different galaxy type models with data is shown in Figure \ref{f:120327}. 
Looking at the $\alpha$-elements, we find that $[Si/Fe]$ is better explained by spiral disk models, even if the extent of the error bars does not allow us to exclude other morphology models. On the other hand, the $[Mg/Fe]$ is well explained by irregular models. The indication of a low SFE galaxy is even stronger from the observed $[S/Fe]$ ratio. In the lower left panel are plotted both the cases in which we consider primary $N$ production by massive stars (\citealt{Matteucci86}) and only secondary production (\citealt{Nomoto13}). In the first case, abundance data are compatible with both the irregular and the spiral reference models.  Also in the case of secondary production both the scenarios remain acceptable, but the agreement is worse . The $[Ni/Fe]$ observed ratio stays between the patterns for irregular and spiral galaxy models (however, we remember the uncertainties in $Ni$ yields). Concerning $Zn$, the observed $[Zn/Fe]$ agrees with the predictions of the models for spirals and spheroids, whereas it remains a little too high for the irregular models. As for GRB 081008 host, a lower dust accretion timescale (i.e faster dust accretion) could explain the abundance ratio in terms of an irregular galaxy. Moreover, the relatively low $[Zn/Fe]$ ratio could be indicative of a late type galaxy, since in hosts with $[\alpha/Fe]$ typical of spheroids we see much higher $[Zn/Fe]$.\\
By means of the statistical test adopted, we find that the best model represents a moderately star forming irregular ($M_{inf}=5\cdot10^9M_\odot$, $\nu=0.2Gyr^{-1}$). The predictions of this model are shown in Figure \ref{f:120327_1}. For the best model we assume lower condensation efficiencies relative to the irregular reference model (i.e. $\delta_{MP}$ instead of $\delta_{HP}$). As a matter of fact, the different behaviour shown by the various elements (in particular $\alpha$ elements) is better reproduced by the $\lq\lq$intermediate scenario" (i.e. $\delta_{MP}$) of \citet{Piovan11}.\\ 
In conclusion, we identify this host galaxy as a moderately massive and star forming irregular galaxy. In this way, we find that the age for this host at the time of the GRB event is $\sim 0.75$ $Gyr$.

\subsubsection{GRB 120815A}     \label{sss:120815A}
In Figure \ref{f:120815} we compare the ranges of the patterns for different galaxy types with the observed abundances for this host. The $[\alpha/Fe]$ ratios ($Si$ and $S$) are very high: this feature is an indicator of an early type galaxy. The observed $[Zn/Fe]$ is also very high and the models are unable to fit it. As for GRB 050820 (see \ref{sss:050820}), however, we note that $Zn$ seems to behave like an $\alpha$ element.
The $[Ni/Fe]$ panel instead would show good agreement with the irregular galaxy models, but we do not consider this element as a discriminant because of its uncertain yields.\\
In Figure \ref{f:120815_1} we report the results of the best model, a star forming spheroid with very high SFE ($25 Gyr^{-1}$) and infall mass ($10^{12}M_\odot$). This model reproduces quite well the observed $[Si/Fe]$, whereas the $[S/Fe]$ is underpredicted. A possible explanation can be found in a higher production of $Fe$ dust from massive stars, or in a lesser depletion of $S$, an element whose tendency to be refractory is still unclear \citep{Jenkins09,Calura09}.\\
For the scenario of the best model, the host age is found to be $\sim10$ $Myr$. 
Such a young age can also explain the mass observed for this host. Looking at Figure \ref{f:mass_host}, in fact, spheroids models give for the stellar mass values similar to the one found by \citet{Kruhler13}, namely $10^{10}M_\odot$. The SFR we predict for this host at the time corresponding to the age of $10 Myr$ and for the stellar mass we obtain (see Figure \ref{f:mass_host}) is $\sim15M_\odot yr^{-1}$, in reasonable agreement with what oserved in this host ($\sim 5 M_\odot yr^{-1}$, given the uncertainties of these measurements, see \citealt{Kruhler15}\footnote{in the references, the SFR is computed using a \citet{Chabrier03} IMF. The value presented here is obtained by converting to a \citet{Salpeter55} IMF (SFR$_{Salpeter}=$ SFR$_{Chabrier}\cdot 1.8$)}, and of galaxy evolution models).

\subsubsection{GRB 161023A}     \label{sss:161023A}

\begin{figure*}
\centering
\includegraphics[width=.81\textwidth]{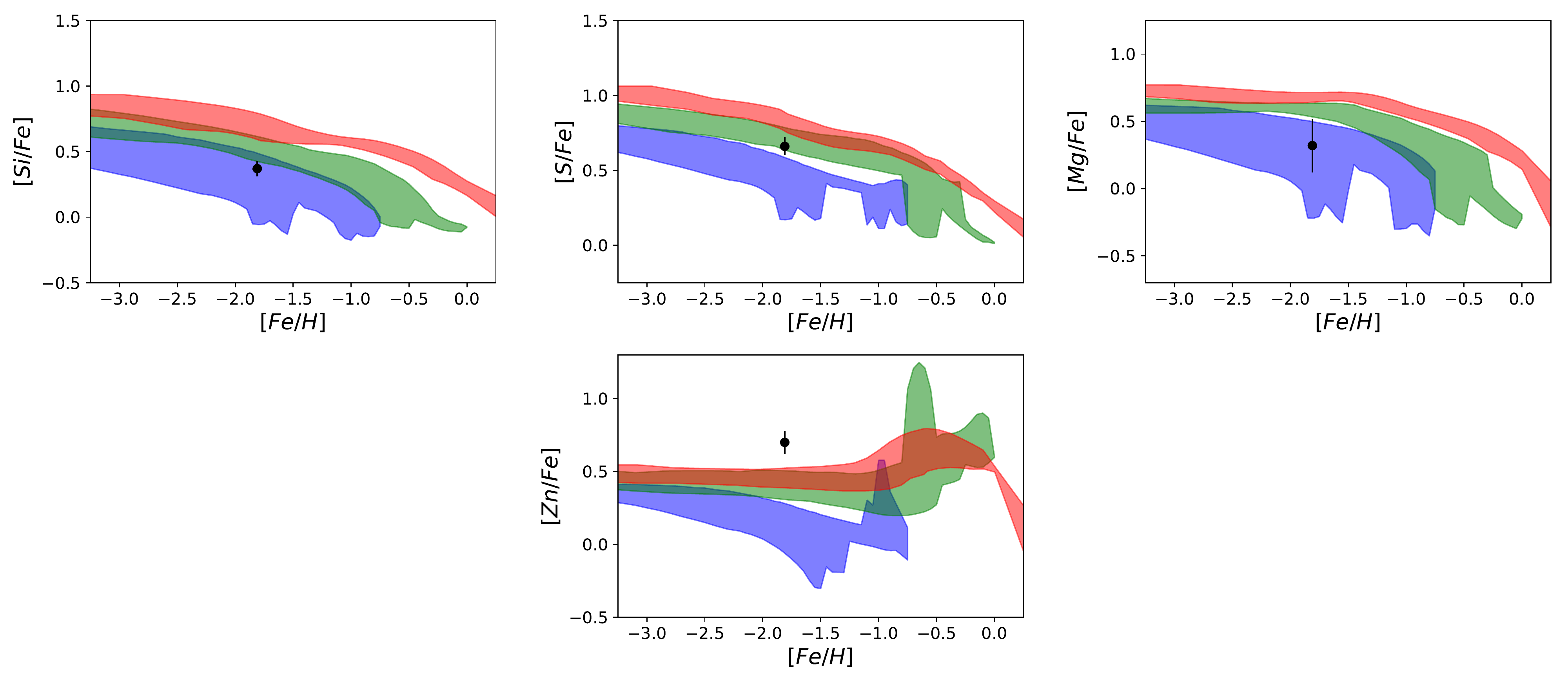}
\centering
\caption{Observed $[X/Fe]$ vs. $[Fe/H]$ ratios for the GRB 161023A host galaxy provided by \citet{deUgarte18}. Data are black symbols with error bars. The blue, green and red shaded areas are the predictions computed by means of models for irregular, spiral and spheroidal galaxies, respectively.}
\label{f:161023}
\end{figure*}

\begin{figure*}
\centering
\includegraphics[width=.81\textwidth]{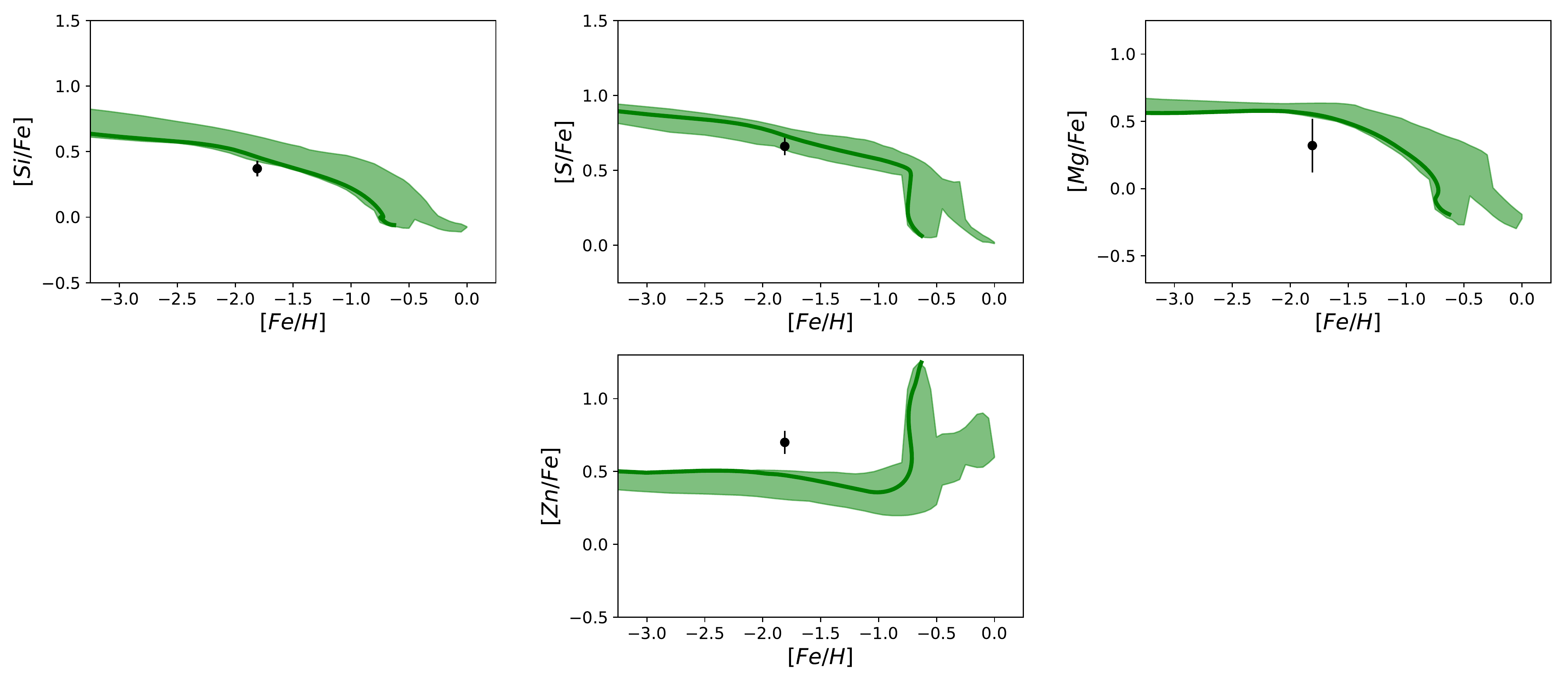}
\caption{Observed $[X/Fe]$ vs. $[Fe/H]$ ratios for the GRB 161023A host galaxy provided by \citet{deUgarte18}. Data are black symbols with error bars.  The green shaded area and the green line are the predictions computed by means of all spiral galaxy models and the best model for the GRB host, respectively.}
\label{f:161023_1}
\end{figure*}

In Figure \ref{f:161023}, the comparison between the models for different galactic type and the data is shown. Here we can see a different behaviour of the observed $[\alpha/Fe]$ relative to the predictions of the models: $[S/Fe]$ is in good agreement with the models for spirals, whereas $Si$ and in particular $Mg$ are better reproduced by the models for irregulars. The observed $[Zn/Fe]$ remains instead above the predictions of the models. \\
In Figure \ref{f:161023_1}, we show the predictions of the best model, a spiral disk one with moderate mass and SFE ($M_{inf}=10^{10}M_\odot$, $\nu=1Gyr^{-1}$). For this host we find that an increased dust production by massive stars (relative to the reference model for spirals, i.e. $\delta_{HP}$ instead of $\delta_{MP}$) helps to explain the observed abundances. \\
For the model with such characteristics, we estimate a host age of $0.15$ $Gyr$.

\newpage
\subsection{IMF Effects}        \label{ss:IMF_effects}

\begin{figure*}
\centering
\includegraphics[width=.81\textwidth]{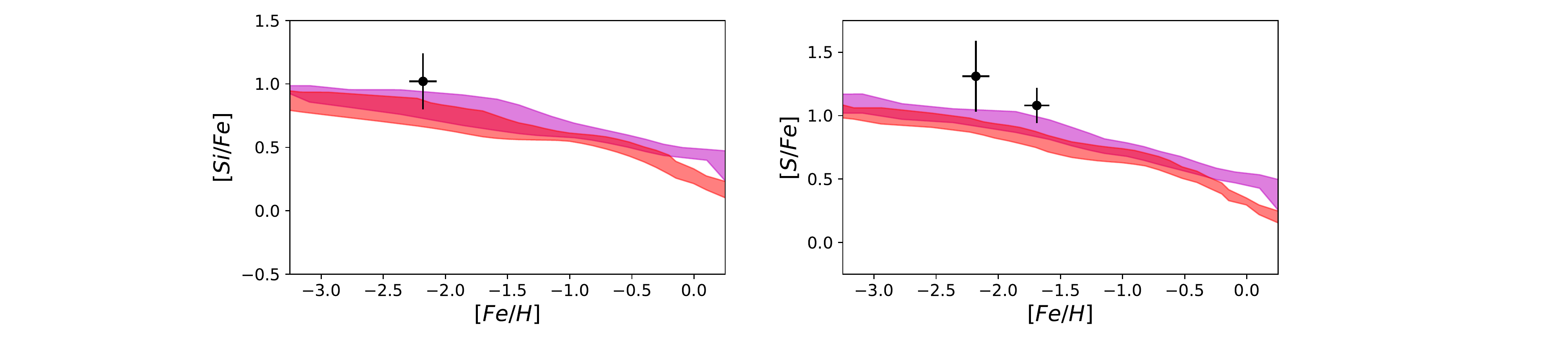}
\caption{Observed $[X/Fe]$ vs. $[Fe/H]$ ratios for the GRB 120815A (left panel) and GRB 050820 (right panel) host galaxies. Data are black symbols with error bars. The red and magenta shaded areas are the predictions computed by means of models for spheroidal galaxies with a \citet{Salpeter55} IMF and a top-heavy IMF, respectively.}
\label{f:top_heavy}
\label{f:IMF_effects}
\end{figure*}

Now we discuss the effects of the IMF in models for different galaxy types. The modification of the IMF, in fact, alters significantly the results given by chemical and dust evolution equations. In our paper we decide to adopt the classical \citet{Salpeter55} IMF, the \citet{Scalo86} IMF and a top heavy one (defined in \ref{ss:birthrate_f}). The differences between these three functions, with respect to chemical enrichment, are mainly due to the slope in the range of massive stars. In the \citet{Scalo86} IMF the number of massive stars is depressed with consequent lower metal enrichment. On the contrary, in the \citet{Salpeter55} IMF (and even more in the top-heavy one) the metallicity is higher and we also expect a larger overabundance in the $[\alpha/Fe]$ ratios. At the same time, the IMF has also an effect on the occurrence of galactic winds, since these are driven by SN explosions. In more recent times, other IMFs have been suggested such as those of \citet{Kroupa01} and  \citet{Chabrier03}. The slope of the IMF of \citet{Kroupa01} is very similar to the \citet{Salpeter55} one ($x=1.3$ against $x=1.35$) and the \citet{Chabrier03} has two versions with one similar to \citet{Salpeter55} and the other to \citet{Scalo86} ($x=1.7$), both in the range of massive stars.\\ 
In this work we test if a change in the IMF can give more clues on the identification of GRB host galaxies. As mentioned in Section \ref{s:chem_model}, computations are also done adopting a \citet{Scalo86} IMF for spiral and irregular models and a top-heavy IMF for the spheroids models.\\
The data from both the identified star forming spheroidal hosts (GRB 050820, GRB 120815A) are better explained by models adopting the top-heavy IMF, as defined in Equation \eqref{e:top_IMF}. In Figure \ref{f:top_heavy} we show what happens in the case of using this IMF or the \citet{Salpeter55} one. From the Figure, it is evident that when adopting the top-heavy IMF we have results more consistent with the data. However, this result is not surprising and is coherent with previous studies on local elliptical galaxies (\citealt{Arimoto86}; \citealt{Gibson97}; \citealt{DeMasi18_1}) and high redshift spheroids \citep{Zhang18}, that claim the adoption of an IMF flatter than the \citet{Salpeter55} one can explain many of their chemical features. A top-heavy IMF also causes a slight $[Zn/Fe]$ enhancement, but not enough to remove the discrepancies seen in the systems studied here. 
Apart from these considerations, the change in the IMF does not bring any significant modification in our conclusions for GRB 050820 and GRB 120815A hosts. In fact, the parameters that better explain the observed abundances remain the same of the best model adopting the \citet{Salpeter55} IMF. At the same time, the predicted ages of the hosts remain of the same order of magnitude ($\sim 10$ $Myr$) of what found before.\\
\begin{figure*}
\centering
\includegraphics[width=.81\textwidth]{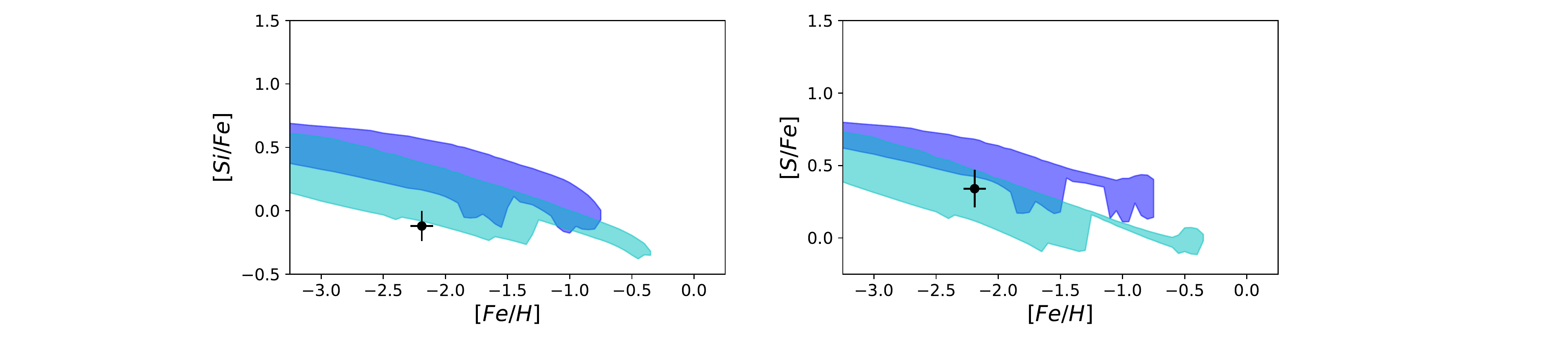}
\caption{Observed $[X/Fe]$ vs. $[Fe/H]$ ratios for the GRB 090926A host galaxy. Data are black symbols with error bars. The blue and the cyan shaded areas are the predictions computed by means of models for irregular galaxies with a \citet{Salpeter55} IMF and a \citet{Scalo86} IMF, respectively.}
\label{f:Scalo_irr}
\end{figure*}
\begin{figure*}
\centering
\includegraphics[width=.81\textwidth]{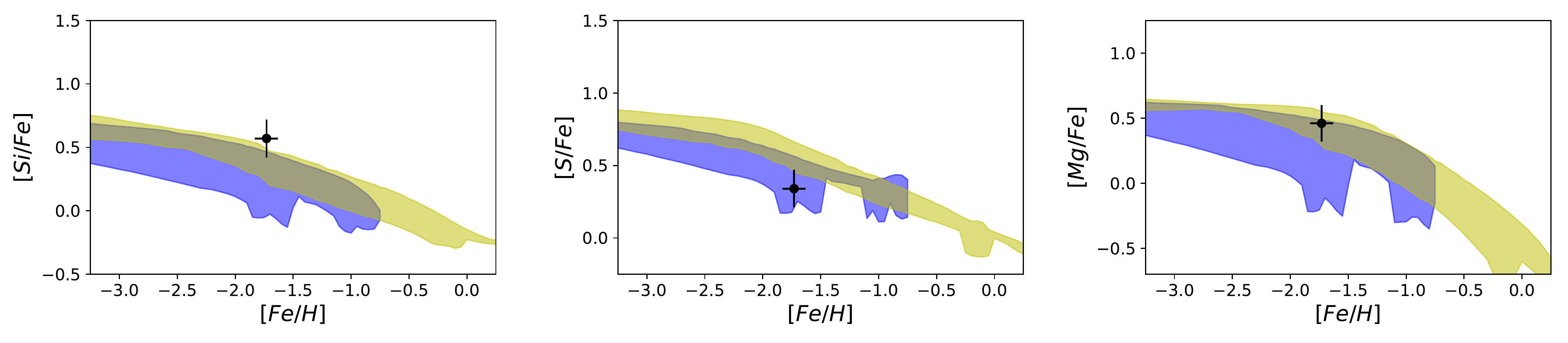}
\caption{Observed $[X/Fe]$ vs. $[Fe/H]$ ratios for the GRB 120327A host galaxy. Data are black symbols with error bars. The blue and the yellow shaded areas are the predictions computed by means of models for irregular galaxies with a \citet{Salpeter55} IMF and spiral disks with a \citet{Scalo86} IMF, respectively. }
\label{f:Scalo_sp}
\end{figure*}
\noindent In the other host galaxies, instead, we test what happens by adopting a \citet{Scalo86} IMF. We find that the adoption of such an IMF in the irregular models for GRB 090926A can better explain the very low $[\alpha/Fe]$ ratios observed. This can be seen in Figure \ref{f:Scalo_irr}. The improved agreement with the observed abundance ratios do not change however the conclusions drawn for GRB 090926A host. Even the age predicted by the best model remain very similar to the one obtained using the \citet{Salpeter55} IMF.\\
Concerning the moderate star forming irregulars GRB 050730 and GRB 120327A hosts, we find that spiral models adopting the \citet{Scalo86} IMF can fit well the observed abundances, as well as the irregular models adopting a \citet{Salpeter55} one. In fact, the statistical test described in Subsection \ref{ss:host_id} give very similar or even better agreement for the \citet{Scalo86} IMF spiral models. In Figure \ref{f:Scalo_sp} we show what happens for GRB 120327A. The partial overlap between the patterns of the irregular, \citet{Salpeter55} IMF models and the spiral, \citet{Scalo86} IMF ones is not surprising. Using a steeper power law for the IMF is equivalent to lower the SFE from the point of view of the abundance patterns. Summarising, the change of the IMF can alter noticeably the results for GRB 050730 and GRB 120327A, affecting the identification of the galaxy type (from irregular to spiral) and hence the age estimated at the time of GRB ($70 Myr$ for GRB 050730 and $0.2 Gyr$ for GRB 120327A as opposed to $0.2 Gyr$ for GRB 050730 and $0.75 Gyr$ for GRB 120327A).\\
The \citet{Scalo86} IMF has to be preferred also to explain the abundances observed in GRB 161023A. Here, however, the change of the IMF does not affects our conclusions on this GRB host, identified as a moderate star forming spiral disk. The data of GRB 081008 instead remain better explained by the spiral disk model adopting a \citet{Salpeter55} IMF. 
The fact that spiral disk models with a \citet{Scalo86} IMF well explain observational data in most of the cases is not unexpected. In fact, this IMF describes better the features of the MW disk than the \citet{Salpeter55} IMF (see \citealt{Chiappini01}; \citealt{Romano05}). On the contrary, for irregular galaxies it was shown that a \citet{Salpeter55} IMF is to be preferrred over a \citet{Scalo86} IMF because it better explains many of their features (see \citealt{Bradamante98}; \citealt{Yin11}).

\section{Summary and conclusions}       \label{s:conclusion}

In this work, we present a method based on detailed chemical evolution models including dust to constrain the nature and the age of LGRB host galaxies. This method was already used in the works of \citet{Calura09} and \citet{Grieco14} and consists in the comparison of the abundance ratios observed in GRB afterglow spectra with abundances predicted for galaxies of different morphological type (irregular, spiral, spheroidal). These are obtained by means of chemical evolution models calibrated on the features of local galaxies. 
The elements considered in this study are $N$, $\alpha$-elements 
, $Fe$-peak elements 
and $Zn$. For the fact that chemical abundances measured from afterglow spectra are ISM gas abundances, the chemical evolution models take into account the dust depletion in the ISM. We adopt in this paper updated and more accurate dust prescriptions with respect to the ones used in previous works. For the stellar yields, we use those which best reproduce the abundance patterns in the solar neighbourhood (see \citealt{Romano10}). Thanks to these improvements, this paper can provide more robust insights on the nature of LGRBs and also on the star formation process in the early universe. As a matter of fact, long GRBs are supposed to be the product of the death of massive stars, and for this reason they can be considered as tracers of the star formation. We analysed the environment of the following 7 GRBs: GRB 050730, GRB 050820, GRB 081008, GRB 090926A, GRB 120327A, GRB 120815A, GRB 161023A.\\
The best models for the various hosts are shown in Table \ref{t:final}, where we report the inferred morphology, galactic age and main parameters (infall mass, infall timescale, SFE, dust prescriptions) adopted for each of the hosts.\\
We summarise our results as follows:
\begin{table*} 
\centering
\tablenum{4}
\caption{Parameters for the best models selected for the hosts.}
\begin{tabular}{c | *{6}c}
\hline
 & Morphology & Age$[Gyr]$ & $M_{inf}[M_\odot]$ & $\tau_{inf} [Gyr]$ & $\nu [Gyr^{-1}]$  & $\delta^{CC}$\\[0.1cm]
\hline
GRB 050730 & Irregular (Spiral) & $\sim0.2$ ($\sim0.07$) & $5\cdot10^{9}$ & $10$ & $0.1$  & $\delta_{MP}$ \\
GRB 050820 & Spheroid & $\sim0.015$ & $10^{12}$ & $0.2$ & $25$ & $\delta_{MP}$  \\ 
GRB 081008 & Spiral & $\sim0.2$ & $10^{11}$ & $4$ & $3$ & $\delta_{HP}$\\
GRB 090926A & Irregular & $\sim1.3$ & $5\cdot10^{8}$ & $11$ & $0.01$ & $\delta_{HP}$\\
GRB 120327A & Irregular (Spiral) & $\sim0.75$ ($\sim0.2$) & $5\cdot10^{9}$ & $9$ & $0.2$ & $\delta_{MP}$\\
GRB 120815A & Spheroid & $\sim0.01$ & $10^{12}$ & $0.2$ & $25$ & $\delta_{MP}$  \\ 
GRB 161023A & Spiral & $\sim0.15$ & $10^{10}$ & $8$ & $1$ & $\delta_{HP}$\\[0.1cm]
\hline
\end{tabular}
\tablecomments{In brackets are reported the morphology and the age from the best model obtained with the \citet{Scalo86} or the top-heavy IMF, when different from the model obtained with a \citet{Salpeter55} IMF.}
\label{t:final}
\end{table*}
\begin{itemize}
    \item The model-data comparison shows that all the three galactic morphological types (irregular, spiral, spheroidal) seem to be present in our sample of host galaxies, confirming the result obtained by \citet{Grieco14} of having also early-type galaxies as hosts of high redshift ($z>2$) GRBs. The high values of $z$ indeed allow us to see early-type galaxies during their period of active star formation, when the massive star explosions are still present and galactic winds are occurring. Such a situation is not found instead at lower redshift, where the star formation is already quenched for this morphological type.
    \item Except for the GRB 090926A, for which the predicted age is higher than $1$ $Gyr$, our models predict ages of the galaxies much younger than a billion year. These very short timescales originate from the low gas metallicities measured in the spectra. As a matter of fact, with the exception of GRB 081008, the gas $[Fe/H]$ values are always $< - 1.5$ $dex$. Considering the whole ISM (gas+dust), this translates into metallicities below $0.1$ the solar value. These low $Z$ values agree with all the main models explaining the origin of LGRBs, where the progenitors of these events should lie (or are at least largely favoured) in a low metallicity environment. 
	\item Concerning $Zn$, we find less good agreement than expected between the observed abundances and the models, especially for the identified early type hosts. Despite our current incomplete understanding of dust depletion, the observed discrepancies seem to be not primarily caused by the dust presence. As a matter of fact, the galactic ages found for these host galaxies are too short to explain the very high observed $[Zn/Fe]$ values in terms of accretion of $Fe$ dust. Moreover, higher dust production by stars (even only for $Fe$) seems to be not the solution. Nevertheless, we note that $Zn$ seems to show an enhancement with respect to $Fe$ at low metallicities, similarly to $\alpha$ elements. This behaviour is found also in MW stars (see \citealt{Romano10}). If this will be confirmed then $Zn$ cannot be taken as a proxy for $Fe$, as instead suggested in the literature \citep{DeCia16}.
    \item  Changing the IMF in the models can help to better explain the abundance ratios observed in some of the host galaxies in the sample. In particular, for GRB 050730 and GRB 120327A we identify an irregular galaxy when using a \citet{Salpeter55} IMF, whereas we identify a spiral when using a \citet{Scalo86} IMF. The fact of having a top-heavy IMF in spheroid models and the \citet{Scalo86} in spiral models agrees with the studies carried out for local and high redshift spheroids (e.g. \citealt{Gibson97}; \citealt{DeMasi18_1}; \citealt{Zhang18}) and the Milky Way (\citealt{Chiappini01}; \citealt{Romano05}). The same is not true for irregular galaxies, where the \citet{Salpeter55} IMF, rather than the \citet{Scalo86}, reproduces many of their features (\citealt{Bradamante98}; \citealt{Yin11}).
    \item We do not always find agreement with the identification results found by \citet{Calura09} and \citet{Grieco14}. In particular, very different is the result for what concerns the GRB 050820, whose host galaxy was identified by \citet{Calura09} as an irregular galaxy with SFE $\nu=0.1 Gyr^{-1}$. In this work we classify this host as a strong star forming ($\nu\ge 15 Gyr^{-1}$) spheroid. Significant differences, but not as strong as for GRB 050820, are also found for GRB 120327A. We ascribe these variations to the adoption of different prescriptions on stellar yields and dust relative to the previous papers. These results strongly highlight the importance of adopting detailed new chemical evolution models with updated and more accurate prescriptions, as it has been done here.
\end{itemize}

\acknowledgments
MP, FM acknowledge financial support from the University of Trieste (FRA2016). FC acknowledges funding from the INAF PRIN-SKA2017 program 1.05.01.88.04.
The authors thank an anonymous referee for careful reading of the manuscript and useful suggestions. 

%

\vspace{5mm}







\bibliography{gal_archeo_GRB}{}
\bibliographystyle{aasjournal}



\end{document}